\documentclass[usenatbib]{mn2e}
\usepackage{epsfig}

\newcommand{\mnras}{MNRAS}
\newcommand{\apj}{ApJ}
\newcommand{\apjl}{ApJL}
\newcommand{\apjs}{ApJS}
\newcommand{\araa}{ARA\&A}
\newcommand{\nat}{Nat}
\newcommand{\sci}{Sci}
\newcommand{\kms}{\hbox{${\rm km\, s^{-1}}$}}
\newcommand{\solarmass}{\hbox{$h^{-1} \rm M_\odot$}}
\newcommand{\solarl}{\hbox{$L_\odot$}}
\newcommand{\kpch}{\hbox{$h^{-1} \rm kpc$}}
\newcommand{\mpch}{\hbox{$h^{-1} \rm  Mpc$}}
\newcommand{\aap}{A\&A}
\def\gsim { \lower .75ex \hbox{$\sim$} \llap{\raise .27ex \hbox{$>$}} }
\def\lsim { \lower .75ex \hbox{$\sim$} \llap{\raise .27ex \hbox{$<$}} }

\title[The earliest stars and their relics in the Milky Way]
{The earliest stars and their relics in the Milky Way}
\author[Gao et al.]{\parbox{18cm}{
L. Gao$^{1,2}$\thanks{Email:lgao@bao.ac.cn}, 
Tom Theuns$^{2,3}$, 
C. S. Frenk$^2$,
A.~Jenkins$^2$, 
J.~C.~Helly$^2$, 
J.~Navarro$^4$, 
V.~Springel$^5$
and S. D. M. White$^5$
}\vspace{0.3cm}\\
$^1$National Astronomical Observationaries, Chinese Academy of Science,
Beijing, 100012, China \\
$^2$Institute of Computational Cosmology, Department of Physics,
University of Durham,\\ Science Laboratories, South Road, Durham DH1
3LE \\
$^3$ Universiteit Antwerpen, Campus Groenenborger, Groenenborgerlaan
171, B-2020 Antwerpen, Belgium\\ 
$^4$ Department of Physics and Astronomy, University of Victoria, PO
Box 3055 STN CSC, BC V8W 3P6, Canada \\
$^5$Max-Planck Institute for Astrophysics, Karl-Schwarzschild Str. 1,
D-85748, Garching, Germany \\
}

\begin{document}
\label{firstpage} \maketitle

\begin{abstract}
We have implemented a simple model to identify the likely sites of
the first stars and galaxies in the high-resolution simulations of
the formation of galactic dark matter halos of the {\em Aquarius
Project}. The first star in a galaxy like the Milky Way formed
around redshift $z=35$; by $z=10$, the young galaxy contained up to
$\sim 3\times 10^4$ dark matter haloes capable of forming stars by
molecular hydrogen cooling.  These minihaloes were strongly
clustered and feedback may have severely limited the actual number
of Population~III stars that formed.  By the present day, the
remnants of the first stars would be strongly concentrated to the
centre of the main halo. If a second generation of long-lived stars
formed near the first (the first star relics), we would expect
to find half of them within 30\kpch\ of the galactic centre and a
significant fraction in satellites where they may be singled out by
their anomalous metallicity patterns. The first halo in which gas
could cool by atomic hydrogen line radiation formed at $z=25$; by
$z=10$, the number of such \lq first galaxies\rq\ had increased to
$\sim 300$. Feedback might have decreased the number of first
galaxies at the time they undergo significant star formation, but not
the number that survive to the present because near neighbours
merge. Half of all the ``first galaxies'' that form before $z=10$
merge with the main halo before $z\sim 3$ and most lose a significant
fraction of their mass. However, today there should still be more than
20 remnants orbiting within the central $\sim 30\kpch$ of the Milky
Way. These satellites have circular velocities of a few kilometers
per second or more, comparable to those of known Milky Way
dwarfs. They are a promising hunting ground for the remnants of the
earliest epoch of star formation. 
\end{abstract}


\section{Introduction}
The first stars probably formed through molecular hydrogen cooling of gas
compressed in dark matter potential wells (Couchman \& Rees 1986;
Tegmark et al. 1997; see Bromm \& Larson 2004 for a recent review). The
temperature of the gas needs to be $\ge 10^3$K for H$_2$ formation to
be efficient; in a $\Lambda$CDM universe dark matter haloes with deep
enough potential wells to heat the gas to this temperature form as
early as redshift $z\sim 50$ (Gao et al. 2004, Reed et al. 2005). Such
stars were very massive, $M\sim 10^2\,M_\odot$ (Abel, Bryan \& Norman
2000; Bromm, Coppi \& Larson 2002; Yoshida et al. 2007; Gao et al. 2007)
and, depending on their initial mass, would have ended their short
lives as supernovae or by collapsing directly into a black hole (Heger
et al. 2003). In neither case can the remnants be identified today.

Very massive first stars (\lq Population~III\rq\ or \lq Pop.~III\rq\ stars)
emit copious amounts of UV photons (Schaerer 2002) which may affect
star formation in nearby haloes through photoevaporation (Shapiro,
Iliev \& Raga 2004; Whalen et al. 2008) and/or destruction of
molecular hydrogen by photons in the Lyman-Werner bands (Haiman, Rees
\& Loeb 1997). These photons could contribute to the
reionization of the Universe (e.g. Sokasian et al. 2004). Elements
synthesised during stellar evolution may pollute the surroundings, 
affecting the cooling rate of gas (Bromm, Yoshida \& Hernquist
2003). The combined \lq feedback\rq\ of photoionisation, energy
injection and changes in composition may allow the formation of lower
mass stars ($M\sim 40M_\odot$, Yoshida et al. 2007) and their feedback
may eventually allow the formation of even lower mass stars, some of
which could still be present today. These very old stars could provide
\lq archaeological\rq\ evidence for early star formation through, for
example, unusual element abundance patterns ({\em e.g.} Iwamoto et al.
2005). As structure formation progresses, more massive dark matter
haloes form in which gas can now cool by radiation from {\em atomic}
hydrogen. This can occur  as early as redshift $z\sim 40$ (Reed et
al. 2005).  Star formation in such \lq first galaxies\rq\ may be very
different from that in the first haloes (see e.g. Wise \& Abel 2008)
because of more efficient cooling by HI and the metals released by
previous Pop.~III stars.\\

There are currently no direct observational constraints on this
scenario and so many uncertainties remain. For example, first star
formation in a Universe where structure on small-scales is suppressed
due to free-streaming of the dark matter (as in Warm Dark Matter
models) is very different (Gao \& Theuns 2007); annihilating dark
matter, not fusion, may power the first stars if the dark matter
consists of Majorana particles (Freese et al. 2008). Whilst the initial
stages of cloud collapse can be understood in terms of the properties
of the ${\rm H}_2$ molecule (Abel et al. 2002; Yoshida et al. 2006),
uncertainties from the modelling of radiative feedback of the
proto-star make the final stellar mass uncertain (McKee \& Tam 2008),
although the general view is that these stars are massive.
The second generation stars may, in fact, have relatively high
metallicity if metal mixing is inefficient. Ionisation front
instabilities (Whalen and Norman 2008) and the explosive growth of
H$_2$ following photoionization may also strongly affect the next
generation of stars (Whalen et al. 2008). \\

A number of important questions remain unanswered. Did star formation
in the Milky Way galaxy start with a display of fireworks? If so,
where should we look for archaeological evidence?  Will most very old
stars be in the bulge (White \& Springel 2000) or will some be found
in the halo? Two recently discovered \lq hyper metal poor\rq\ stars
($[\rm Fe/H]<-5$) have peculiar element abundances. Were these
second-generation stars enriched by the first stars\footnote{We will
  refer to nearly primordial stars contaminated only by metals from
  the first stars as \lq first star relics\rq.}  (see the review by
Beers \& Christlieb 2005)? Finding more such unusual stars in the
Milky Way is a key science driver for new surveys such as {\small
  SEGUE}, and {\small LAMOST}. Did any of the first galaxies survive
to the present day? Where are their remnants? Are they related to the
newly discovered dwarfs (Willman et al. 2005; Belokurov et al. 2006),
as discussed in a number of recent papers (Madau, Diemand \& Kuhlen
2008; Bovill \& Ricotti 2009; Koposov et al. 2009). If so, what
constraints do these place on the earliest star formation?  Helmi et
al. (2006) argued that the remnants of early star formation in the
Milky Way differ from those in dwarf galaxies because the latter had
already been polluted with metals. However, Kirby et al. (2008) drew
the opposite conclusion from more detailed modelling of stellar
metallicities.\\

In this paper we address some of these questions using a suite of
ultra-high resolution cosmological dark matter simulations of galactic
haloes -- the {\em Aquarius Project}. The six dark matter haloes data
from the {\em Aquarius Project} picked from a large cosmological
simulation to be similar in mass to that of the Milky Way, and they
were simulated at a variety of numerical resolutions. This enables us to
investigate halo-to-halo variations, and establish the limitations
imposed by resolution. We employ simple models to identify the sites
of first star or first galaxy formation and follow a suitable
selection of dark matter particles to track where their progenitors
are today, yielding generic cold dark matter predictions for early
star formation in galaxies like the Milky Way.\\

The outline of the paper is as follows. In Sections 2, 3 and~4
respectively we describe the simulations, our models for identifying
first star and first galaxy haloes, and investigate the clustering of
these haloes. In Section~5 we trace the early objects to $z=0$ and
discuss their properties and spatial distribution. Finally, in
Section~6 we present a summary and discussion of our results.

\section{Simulations}
The dark matter simulations analyzed in this study come from the {\em
Aquarius Project}\footnote{http://www.mpa-garching.mpg.de/aquarius/}
(Springel et al. 2008a, b; Navarro et al. 2009): a set of six haloes
similar to that of the Milky Way (virial mass $M_{\rm 200}\sim
10^{12}$\solarmass, circular velocity $V_{\rm 200} \sim
180$~km~s$^{-1}$, where the subcript 200 refers to the region within
which the density is 200 times the cosmic critical density; see
Table~\ref{TabSims}). The halos were identified in a larger
cosmological simulation, the {\small hMS} simulation (Gao et
al. 2008), which follows the growth of structure in a periodic box of
comoving length 100\mpch\ with $900^3$ particles, assuming the same
cosmological parameters as in the {Millennium
  simulation\footnote{http://www.mpa-garching.mpg.de/galform/virgo/millennium/}}
(Springel et al. 2005): $\Omega_m=0.25,\Omega_\Lambda=0.75, h=0.73,
\sigma_8=0.9$. The {\small hMS} is itself a lower resolution version
of the Millennium-II
{simulations\footnote{http://www.mpa-garching.mpg.de/galform/millennium-II/}
 (Boylan-Kolchin et al. 2009), so that the {\em Aquarius} haloes are
 alos present in the Millennium-II simulation. 

Resimulating the formation of haloes in such a large box correctly
accounts for tidal forces and yields realistic formation times. The
\lq zoom\rq\ technique for resimulating haloes represents the mass
which ends up inside or near the chosen halo with many low mass
particles and a small gravitational softening; the mass in more
distant regions is represented with increasingly more massive
particles with larger gravitational softening. Appropriate small-scale
power is added in the initial conditions to perturb the high-resolution
particles. The haloes, labelled Aq-A to Aq-F, are simulated with up to
$10^9$ particles within the virial radius (particle mass $\sim
10^{3-4}$\solarmass), using the Gadget-3 simulation code which is an
updated version of Gadget-2 (Springel 2005). The Aq-A halo has been
resimulated at different resolutions in order to study
numerical convergence.\\

\begin{table*}
\begin{tabular}{lccrrccccr}
\hline 
Name & $m_{\rm p} [h^{-1}{\rm M_{\odot}}]$ & $\epsilon [h^{-1}{\rm
  pc}]$ & $M_{\rm 200} [h^{-1} {\rm M_{\odot}}]$ & $r_{\rm 200} [h^{-1} {\rm kpc}]$ &
$V_{\rm max} [{\rm s^{-1}km}]$ & $V_{\rm 200} [{\rm s^{-1}km}]$\\
\hline 
Aq-A-1   &  $1.25 \times 10^3$  & 15.0   & $1.34\times 10^{12}$ &
179.40 & 208.8 & 179.2\\
Aq-A-2   &  $1.00 \times 10^4$  & 48.0   &$1.34 \times 10^{12}$ &
179.40  &208.5 & 179.2\\
\hline
Aq-B-2   &  $4.70\times 10^3$  & 48.0    &$0.60 \times 10^{12}$ &
136.51 &157.7 &137.5\\
\hline
Aq-C-2   &  $1.02\times 10^4$  & 48.0   &$1.30\times 10^{12}$ & 177.26
& 222.4 & 177.6\\
\hline
Aq-D-2   &  $1.02\times 10^4$  & 48.0   &$1.30\times 10^{12}$ & 177.26
& 203.2 &177.6\\
\hline
Aq-E-2   &  $7.00\times 10^3$  & 48.0   &$0.87 \times 10^{12}$ &
155.00 & 179.0 & 155.4\\
\hline
Aq-F-2   &  $4.95\times 10^3$  & 48.0   &$0.83\times 10^{12}$ & 152.72
& 169.1 &153.2
\\

\hline
\end{tabular}
\caption{Properties of the six {\em Aquarius} haloes, Aq-A -- Aq-F. The Aq-A
  halo has been resimulated at a variety of numerical resolutions,
  labelled Aq-A-1 -- Aq-A-5. (In this paper, we analyze only Aq-A-1
  and Aq-A-2.)  From left to right the columns give: the halo identifier,
  particle mass, (comoving) force softening length in the
  high-resolution region, virial mass, virial radius, and maximum and
  virial circular velocity, all at redshift $z=0$ ($M_{200},
  \,r_{200}, \,V_{\rm max}$ and $V_{\rm 200}$, respectively).}
  \label{TabSims}
\end{table*} 

Snapshots are stored at 128 output times equally spaced in $\log(a)$,
where $a=1/(1+z)$ is the expansion factor, between redshifts $z=127$
and $z=0$. Dark matter haloes and their substructure are identified
using a combination of the friends-of-friends algorithm (Davis et
al. 1985) and {\small SUBFIND} (Springel et al. 2001).  {\small
SUBFIND} groups particles into substructures based on their density
and binding energy. Haloes and their substructure are traced through
the snapshots and linked together in a merger tree, as in Cole et
al. (2008).\\

Although the {\em Aquarius} haloes range in mass only from $0.6 \times
10^{12}$\solarmass\ to $1.3 \times 10^{12}$\solarmass,
their mass accretion histories, plotted in
Fig.~\ref{fig:massaccretion}, differ significantly.  For example, the
mass of the main progenitor of the Aq-F halo is nearly an order of
magnitude lower than that of Aq-C as late as $z \sim 1$, and similarly
large differences occur for other haloes at higher $z$.  The thick
dashed line is the extended Press-Schechter (Press \& Schechter 1974;
Bond et al. 1991; Lacey \& Cole 1993) prediction for a halo of the
median mass, $M = 1.3 \times 10^{12}h^{-1}{\rm M_\odot}$, computed
following Gao et al. (2004a). Apart from a systematic overestimate of
the mass at high redshift ($z> 20$), the prediction works well. 
\begin{figure}
\resizebox{9cm}{!}{\includegraphics{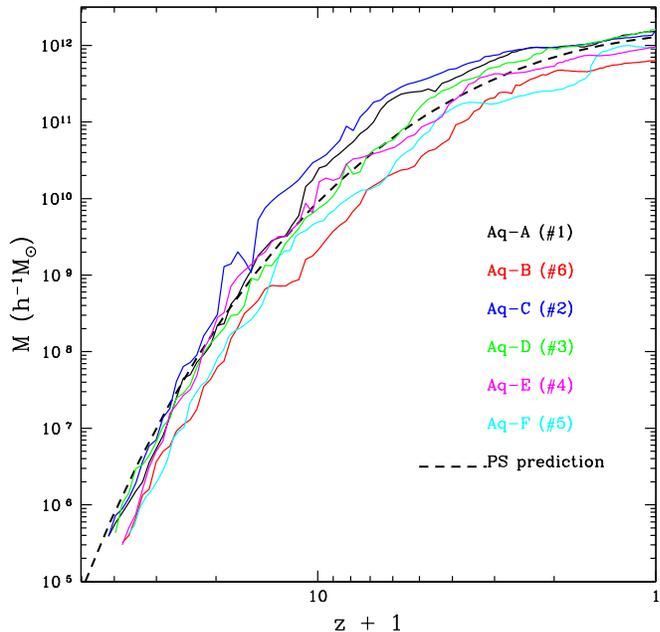}} \\
\caption{Mass of the most massive progenitor of each of the six
  {\em Aquarius} haloes as a function of redshift ($M(z)$, coloured lines),
  compared to the extended Press-Schechter (ePS) theory for a halo of
  mass equal to the median for the six halos, $M = 1.3 \times
  10^{12}h^{-1}{\rm M_\odot}$ (dashed line). There is over an order of
  magnitude variation in $M(z)$ between the haloes at certain
  redshifts, even though the $z=0$ masses are all within a factor of
  two (The number in brackets sorts haloes by their $z=0$ mass.). The ePS prediction works well, but systematically overestimates
  the mass above $z\sim 20$. }
\label{fig:massaccretion}
\end{figure}

\section{Identifying sites of high-redshift star formation}

In this section, we describe a simple model for identifying those
haloes in the simulations which are likely to be the sites of early
star formation.

\subsection{Sites of Pop.~III star formation}
Our model is based on the results of a suite of hydrodynamical
simulations of the formation of early structures in a $\Lambda$CDM
cosmology, carried out by Gao et al. (2007), which included
nine-species, non-equilibrium gas-phase chemistry of ${\rm H}_2$
production. The simulations used zoomed initial condition to follow 
the collapse to high densities ($n_{H} \sim 10^{10}{\rm cm}^{-3}$)
of gas in haloes with a wide range of formation redshifts ($50>z
>10$) and masses ($10^{5}h^{-1}{\rm M_\odot} < M < \hbox{\rm a few}
\times 10^6h^{-1}M_\odot$), and investigated how this depends on the
assumed cosmological parameters. In agreement with previous studies
(e.g. Abel et al. 2002; Bromm et al. 2002; Yoshida et al. 2006), these
authors found that the virial temperature, $T_{\rm vir}$, of the
parent dark matter halo is the key physical parameter that regulates
how efficiently gas collapses and cools. 

Once $T_{\rm vir}\ge 10^3$~K, the formation of molecular hydrogen
becomes very efficient, a significant molecular hydrogen abundance
(${\rm f_{H_2} \sim 10^{-4}}$) builds up and the gas cools rapidly,
becomes self-gravitating and collapses to higher and higher
densities, presumably culminating in the formation of a star. The
duration of this process scales with the dynamical time of the halo
and is typically $\Delta z \sim 3-4$; during this time, a halo may
increase its mass significantly through mergers and
accretions. Therefore neither the instantaneous halo mass, nor the
virial temperature, are sufficient to describe the star forming
properties of a halo; its merger history needs to be taken into
account.\\

Gao et al. (2007) developed a simple model of Pop.~III star
formation, based on the merger trees of the parent halos, which
quite accurately matches the Pop.~III star formation properties of
all 8 individual objects simulated in that study. Define the
  virial temperature $T_{\rm vir}$ of a halo in terms of its circular
  velocity, $V_{200}$, at a radius $R_{200}$ (within which the mean
  density equals 200 times the critical density) as
\begin{equation}
k_{\rm B}\,T_{\rm vir}\equiv {1\over 2}\,\mu\,m_{\rm p}\,V^2_{200}\,,
\end{equation}
where $k_{\rm B}$ is Boltzmann's constant, $\mu$ is the mean molecular
weight ($\mu=1.22$ for neutral H-He admixture), and $m_{\rm p}$ is the
hydrogen mass ({\rm e.g.} White \& Frenk 1991)}. Once the virial
temperature reaches $T_{\rm min}=1100$~K (or, equivalently, the
circular velocity exceeds $V_{200, {\rm min}}=4$\kms), a star is
assumed to form after a redshift delay of $\Delta z \sim 3-4$. We
apply this model to identify which haloes form Pop.~III stars, and
when.\\ 

The first star may suppress or enhance star formation in neighbouring
halos through the combined effect of its ionising and Lyman-Werner
radiation (e.g. Haaiman, Rees \& Loeb 1996; Machacek et al. 2003;
Yoshida et al. 2003, Wise \& Abel 2008; Whalen et al. 2008). The net effect
is still unclear and so we introduce below a simplified feedback model
based on the spatial clustering of the haloes.\\

\subsection{Sites of the first galaxies}
Once the virial temperature of the halo is high enough for atomic line
cooling of gas to become important, $T_{\rm vir} \sim 6000$~K, we
assume that the nature of star formation changes. Rather than a
single, massive primordial star per halo, the cooling of the
metal-enriched gas is assumed to lead to the formation of a normal
stellar population with a range of masses. The depth of the potential
well of these \lq first galaxy\rq\ haloes is still so shallow that
even a single supernova explosion can produce strong winds, as in the
classic White \& Rees (1978) picture, leading to bursty behaviour
which is difficult to model; see e.g. Mori, Ferrara \& Madau (2002);
Yoshida et al. (2007); Wise \& Abel (2008); Greif et al. (2008); Bromm
et al. (2009).  Whether it is simply the contribution of atomic lines
to cooling that leads to the formation of lower mass stars alongside
massive ones, or that of metals or dust (Schneider et al 2006), or the
change in the turbulent properties of the star-forming gas (Krumholtz
\& McKee 2005), or all of the above, is presently unclear. Given these
uncertainties, we adopt the simple criterion, $T_{\rm vir}\ge 10^4$~K,
to separate haloes that form Pop.~III stars from \lq first galaxy\rq\
halos that form stars with a broader IMF. Of course, if feedback from
one halo strongly affects star formation in its surroundings, this
distinction may be too simplistic.

\section{Formation history and clustering of star forming haloes}

The general picture of early star formation sketched above changes
dramatically as soon as the star-forming gas becomes ionised, either
through photons emitted by nearby stars, or because the universe as a
whole becomes ionised.  The photo-heating of the gas and the
suppression of atomic cooling quenches star formation in dwarf
galaxies with circular velocity, $V_{\rm max}\le 25$~km~s$^{-1}$
(e.g. Efstathiou 1992; Thoul \& Weinberg 1996; Gnedin 2000; Hoeft et
al. 2006; Okamoto, Gao \& Theuns 2008).  The Thomson scattering
optical depth towards the last scattering surface, as determined from
the {\small WMAP} data (Komatsu et al. 2008), suggests that the
universe was reionized at a redshift, $z_r\sim 10$. We will consider
earlier times, $z>z_r$, when the Universe was still mainly neutral,
and star formation could take place in haloes with circular velocity
as small as a few kilometers per second.

\subsection{The first stars in galaxies like the Milky Way}
\begin{figure*}
\resizebox{8cm}{!}{\includegraphics{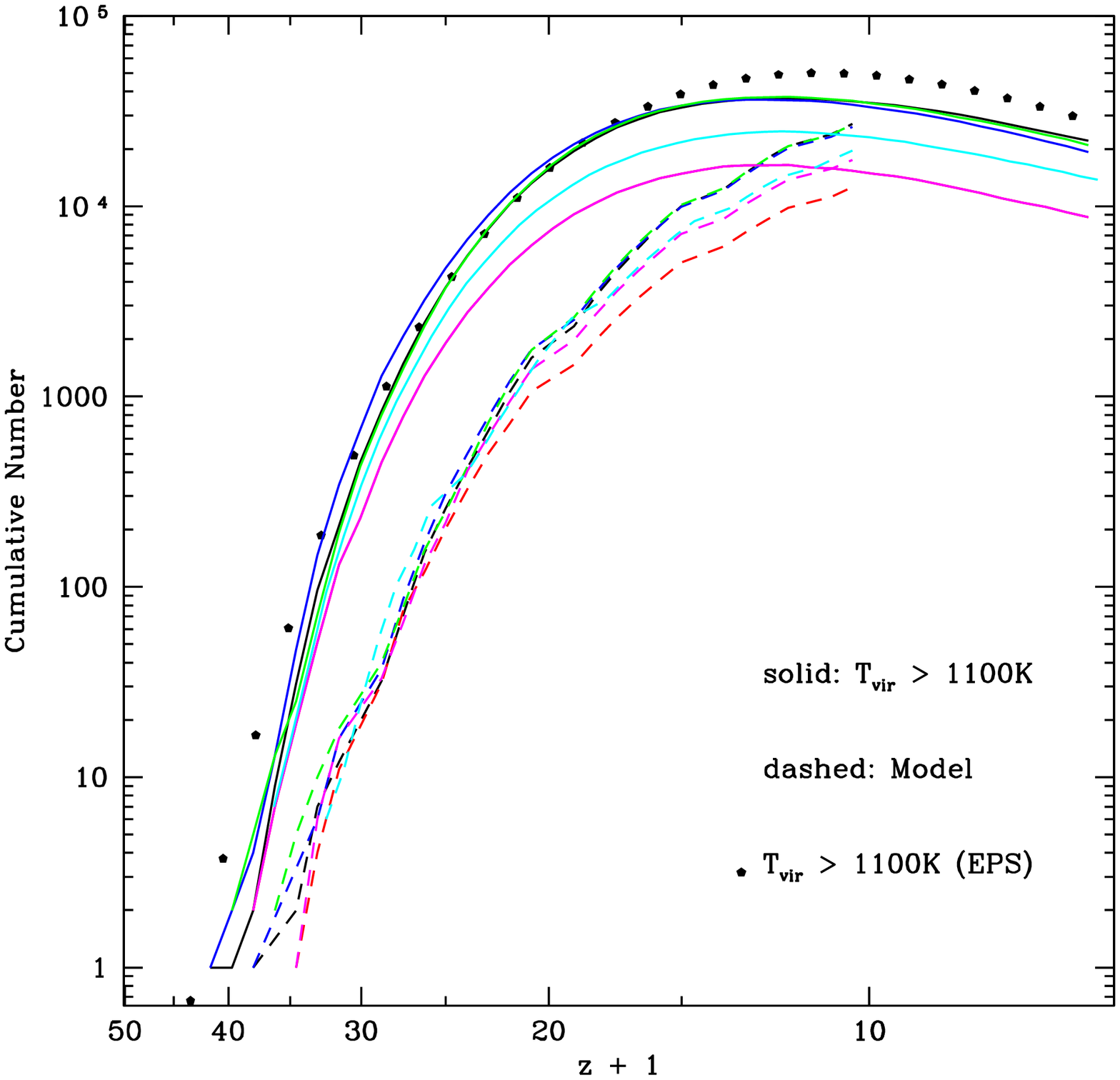}}
\resizebox{8cm}{!}{\includegraphics{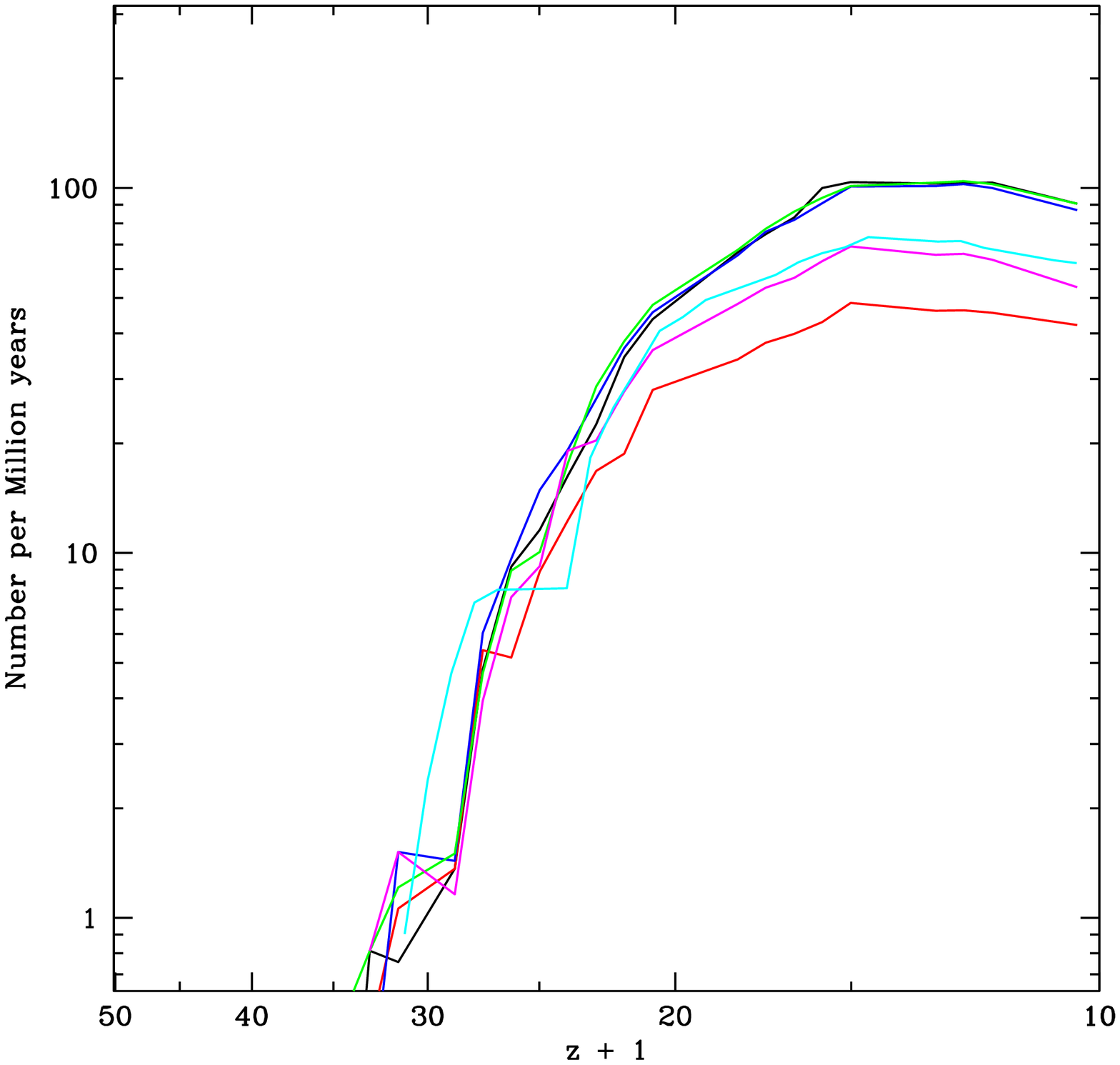}}
\caption{Pop.~III star formation in galaxy halo progenitors 
as a function of redshift. {\em Left panel}: cumulative number of
progenitor haloes with virial temperature above 1100~K for each of the
six {\em Aquarius} haloes (colour solid lines). The black dots show
the extended Press-Schechter prediction which works well, but
increasingly overestimates the number of halos after $z\sim 15$. Gas
in such haloes is hot enough for significant ${\rm H}_2$
formation. The dashed lines show the cumulative number of independent
sites of Pop.~III star formation according to the model of Gao et
al. (2007) which, in addition to a virial temperature criterion, takes
the merger history of the halo into account. This more stringent
criterion reduces the number of forming stars; yet a Milky
Way-type halo still had several hundreds of Pop.~III star-forming
progenitors at redshift $z=25$ and $\sim 10^4$ at $z=10$. The
halo-to-halo variation is large, a factor $\sim 5$ at $z=10$. The
cumulative number of haloes drops after $z\sim 10$ as haloes begin to
merge. {\em Right panel}: the star formation rate. The rate increases
from a few star forming events per Myr at $z=30$, to several tens at
$z=20$ and changes little thereafter.}
\label{fig:starabundance}
\end{figure*}

\begin{figure*}
\resizebox{\columnwidth}{!}{\includegraphics{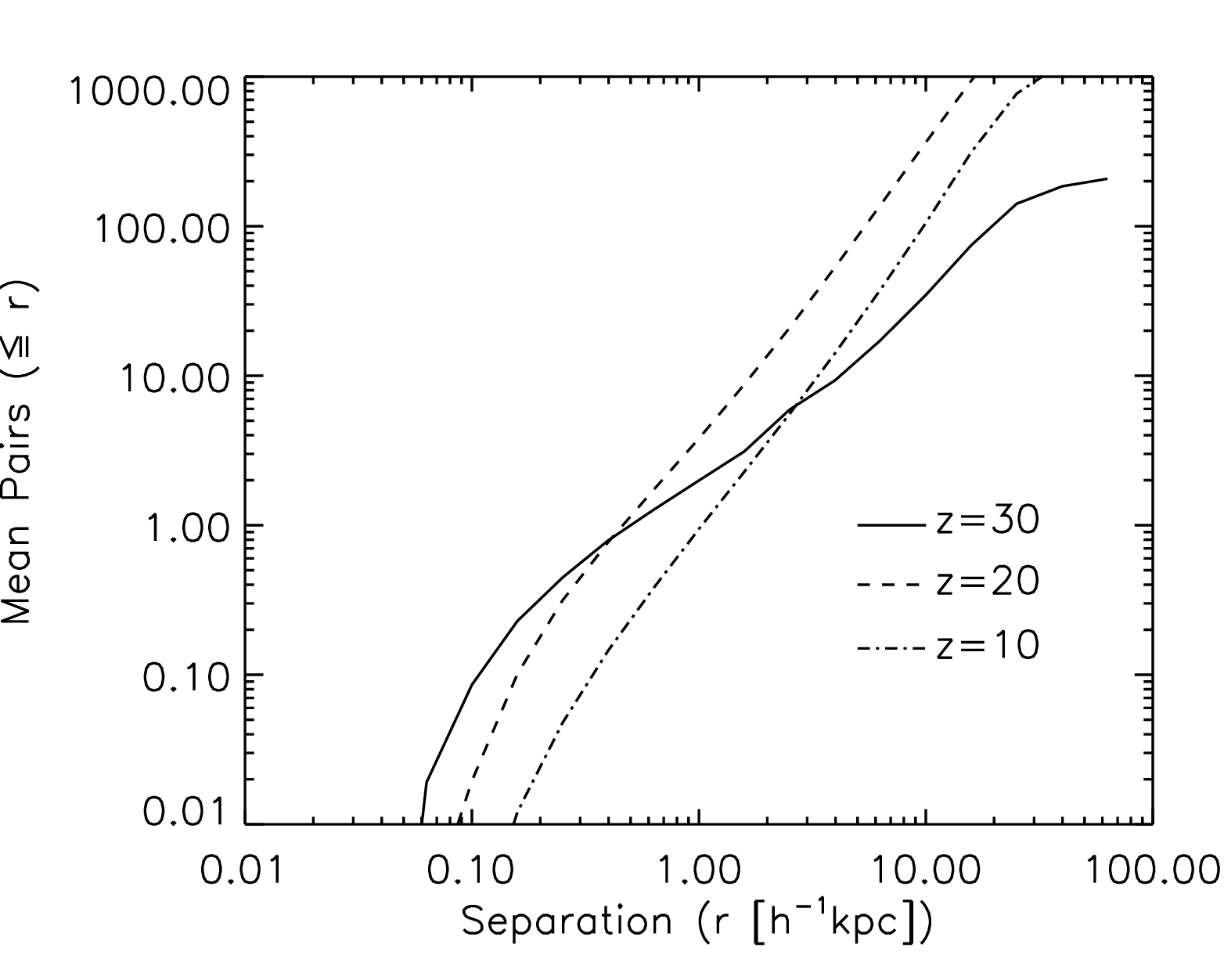}}
\resizebox{\columnwidth}{!}{\includegraphics{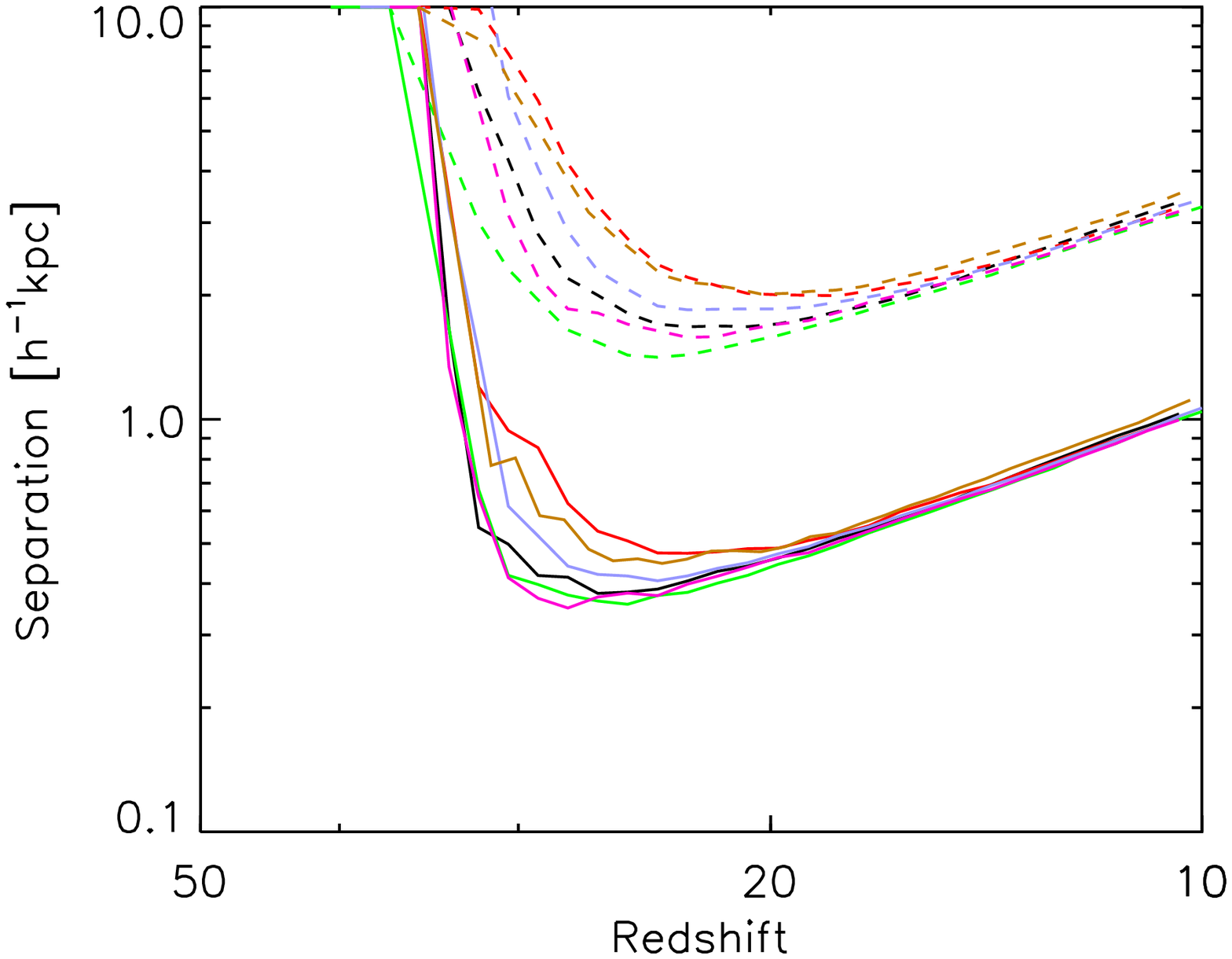}}
\caption{Clustering of  Pop.~III star-forming progenitor haloes.
{\em Left panel:} mean number of progenitor halo pairs in Aq-A, as a
function of proper separation at the three redshifts indicated in the
panel. {\em Right panel}: mean separation of progenitor haloes in each
of the six {\em Aquarius} haloes. Each halo has one neighbour within
the distance given by the solid line, and 10 neighbours within the
distance given by the dashed lines.  Star-forming halos have
relatively nearby companions, indicating that feedback effects could
be important (see text). The linear rise in separation between substructures below $z\sim 20$ is  because the proto-halo region is still expanding with the Hubble flow at this early time.}
\label{fig:starcor2}
\end{figure*}

Figure~\ref{fig:starabundance} shows the evolution in the number of
{\em Aquarius} progenitor dark matter haloes that, in the absence of
feedback, would be able to form Pop.~III stars. The numbers are
calculated either using a simple virial temperature criterion (full
lines) or the more detailed model of Gao et al. (2007; dashed
lines). Star formation begins around $z\sim 40$ ($z\sim 35$ in the Gao
et al. 2007 model), and the number of star-forming progenitor haloes
increases very rapidly to reach $10^3$ at $z\sim 30$ and $10^4$ at
$z\sim 15$; thereafter the numbers decrease as the haloes begin to
merge. The halo-to-halo variation remains relatively small up to
$z\sim 20$ but, by $z\sim 10$, it has increased to about a factor of
five, with more massive {\em Aquarius} haloes tending to have more
star forming progenitors at $z\sim 10$. \\

The difference in the number of star-forming haloes predicted by the
simple, $T_{\rm vir}=1100$~K, cut and the Gao et al. (2007) model is
large, of order a factor 10 or more at $z\ge 15$. The reason for this
difference is the delay (by $\Delta z\approx 3$) in the onset of star
formation in the Gao et al. model at a time when the number of
star-forming objects is rising extremely rapidly. Once the number of
star-forming haloes reaches its maximum, the two models agree to
within a factor of a few. The star formation rate is of order $\dot
N_\star\approx 1\,{\rm Myr}^{-1}$ at $z=30$, increasing rapidly to
reach a broad peak of $\dot N_\star\approx 10^2\,{\rm Myr}^{-1}$ at
$z\sim 15$. According to the Gao et al. model, the first stars in a
Milky Way sized halo appear around $z\sim 35$,  later than in a
proto-cluster where star formation starts around $z\sim 47$ (Gao et
al. 2007). \\

So far we have neglected potential feedback effects, arising from star
formation, on neighbouring haloes which could, in fact, be quite
severe. UV photons with energy, $h\nu> 13.6$~eV, emitted by massive
stars may ionise and heat the gas in a neighbouring halo. Star
formation may thus be quenched due to the suppression of atomic
cooling and photo-heating, or altogether prevented as the halo loses
gas through photo-evaporation (e.g. Shapiro, Iliev \& Raga 2004;
Whalen et al. 2008, Okamoto et al. 2009). On the other hand, H$_2$
formation may be enhanced in photo-ionised gas, once the source of
ionising photons has faded (Haiman, Rees \& Loeb 1996). These effects
will be mostly restricted to the HII region surrounding the star because of the
high optical depth of ionising radiation, except for the high-energy
tail of the photon energy distribution.

Lyman-Werner photons with $11.2\le h\nu/{\rm eV}\le 13.6$ may destroy
the H$_2$ coolant, also quenching or suppressing star formation. Most
of the gas is optically thin to Lyman-Werner photons and so this \lq
sphere of influence\rq\ extends beyond the edge of the HII
region. However, some star-forming haloes may become dense enough to
self-shield, even in the Lyman-Werner bands. Kinematical feedback from
the star in the form of a stellar wind or supernova explosion, may
also affect its surroundings. 

Clearly, the details of these feedback effects are difficult to model
accurately, and so the degree of suppression of star formation near a
Pop.~III star remains highly uncertain (see Yoshida et al. 2003;
O'Shea \& Norman 2007 and 2008; Wise \& Abel 2008; Yoshida et
al. 2008; Whalen et al. 2008; Ahn \& Shaprio 2006, for recent
discussions). In view of this, we have developed a simple model for
feedback that takes into account the clustering of the halos.  We
consider only radiative feedback and neglect external radiation from
distant stars or galaxies. For each progenitor halo, we calculate the
average number, $N(r)$, of neighbouring potentially star-forming
haloes a function of separation, $r$. The number of neighbours, shown
in Fig.~\ref{fig:starcor2}, scales roughly as $N(r)\sim r^2$. At
$z=30$, the typical pair separation is $r\sim 1$\kpch (note that here
and below we give seperations in proper, not comoving units).

A more direct measure of the redshift evolution of clustering is the
mean separation of pairs, $N(1)$, and the mean distance to the 10th
star-forming nearest neighbour halo, $N(9)$, as a function of
redshift, also shown in Fig.~\ref{fig:starcor2}. Both curves fall
rapidly with decreasing redshift as haloes begin to form in
profusion, reach a minimum between $z=30-20$ and thereafter increasing slowly
due to Hubble expansion of the proto-galactic region. On average, each
potentially star-forming halo has another potentially star-forming
halo within 0.5\kpch\ at $z=30$, increasing slowly to 0.6\kpch\ at
$z=20$, and ten others within 2\kpch\ over the same redshift range. 

To estimate the effect of feedback from ionising radiation, we 
compare these numbers to the radius, $R$, of an HII region in a
uniform density medium,
\begin{equation}
R = 1.3\,\left({\dot N_\gamma\over 10^{50}\,{\rm s}^{-1}} \,{\Delta
t\over 3{\rm Myr}}\,{100\over 1+\delta}\right)^{1/3}\,{10\over
1+z}\,{\rm kpc}\,, 
\end{equation}
around a source emitting ionising photons at a rate, $\dot N_\gamma$,
over a time interval, $\Delta t$. The medium is at an overdensity,
$\delta$, has the cosmological baryon fraction, $\Omega_b h^2=0.045$,
and a hydrogen abundance of 0.75 by mass. Full numerical simulations
find similar radii for the HII region (e.g. Abel et al. 2007, Yoshida
et al. 2007). The fact that $R$ is comparable to the mean interhalo
separation implies that feedback from ionising photons could indeed be
rather important in quenching first star formation. The formation rate
of first stars could then be significantly smaller than the no-feedback
case depicted in Fig.~\ref{fig:starabundance}.

\subsection{The first galaxy haloes}
\label{sect:firstgals}
\begin{figure}
\resizebox{8cm}{!}{\includegraphics{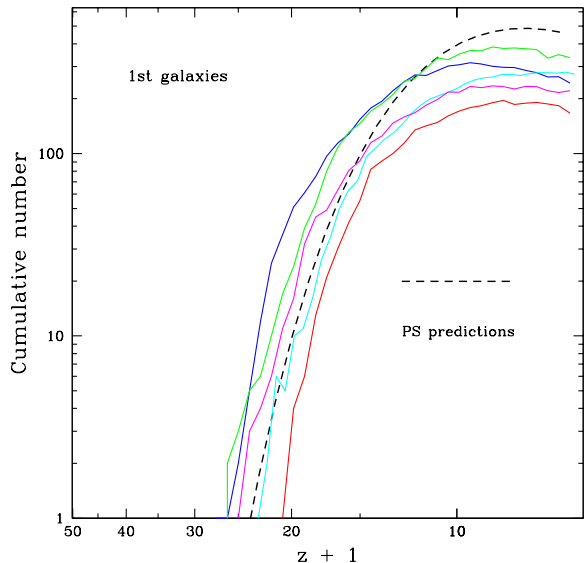}}
\caption{The abundance of haloes hosting \lq first galaxies\rq\
  (virial temperature $T_{\rm vir}>10^4$~K) as a function of redshift
  for the six {\em Aquarius} haloes (colours are the same as in Fig.~
  1). Dashed lines are the prediction from  extended Press-Schechter
  theory. These haloes start to appear around $z\sim 25$ and their
  abundance rises rapidly to reach a broad peak of 200-300 haloes by
  $z\sim 10$. Progenitors of haloes more massive at $z=0$ tend to have
  a higher abundance of first galaxy haloes at all $z$.}
\label{fig:firstgalaxyabundance}
\end{figure}

\begin{figure*}
\resizebox{8cm}{!}{\includegraphics{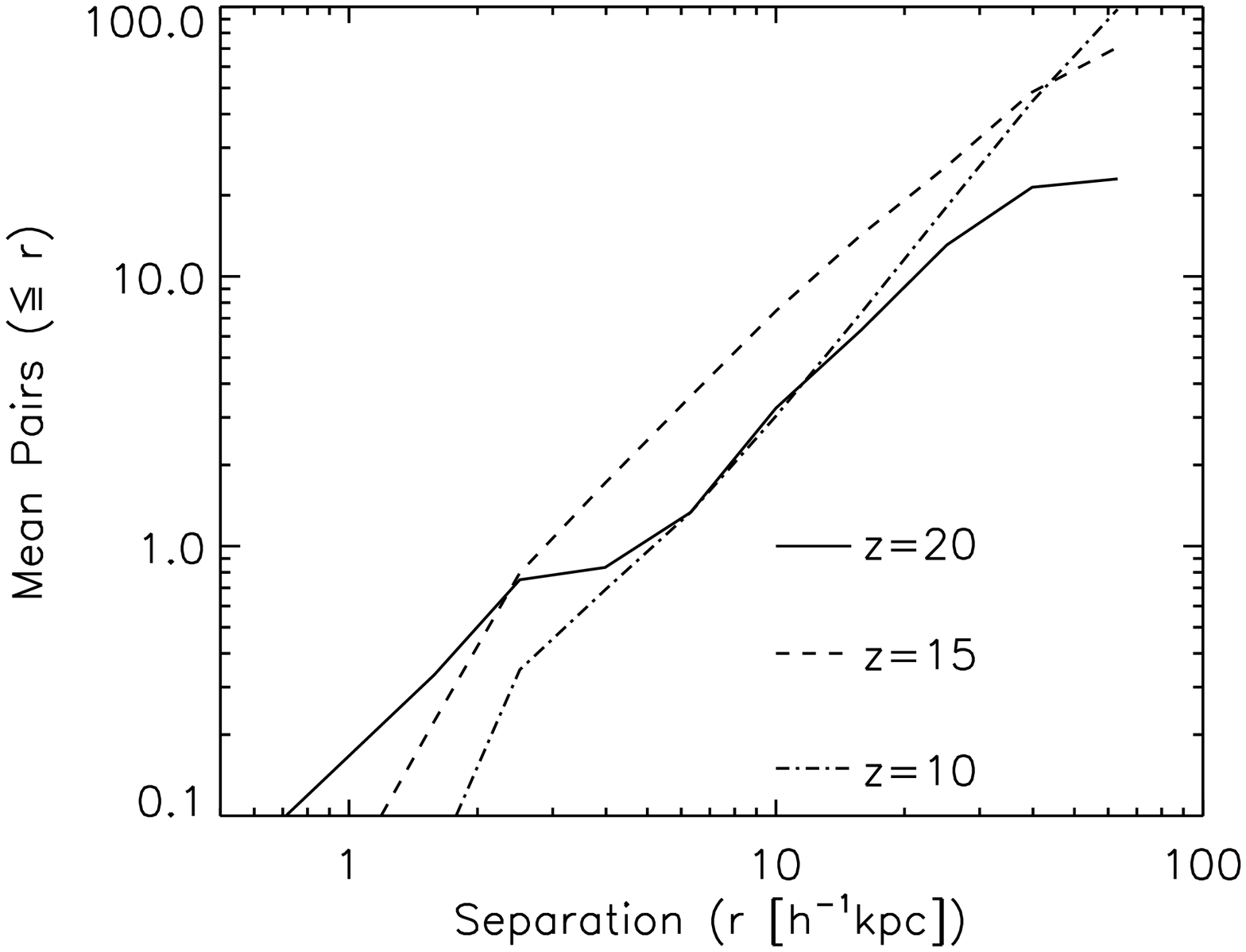}}
\resizebox{8cm}{!}{\includegraphics{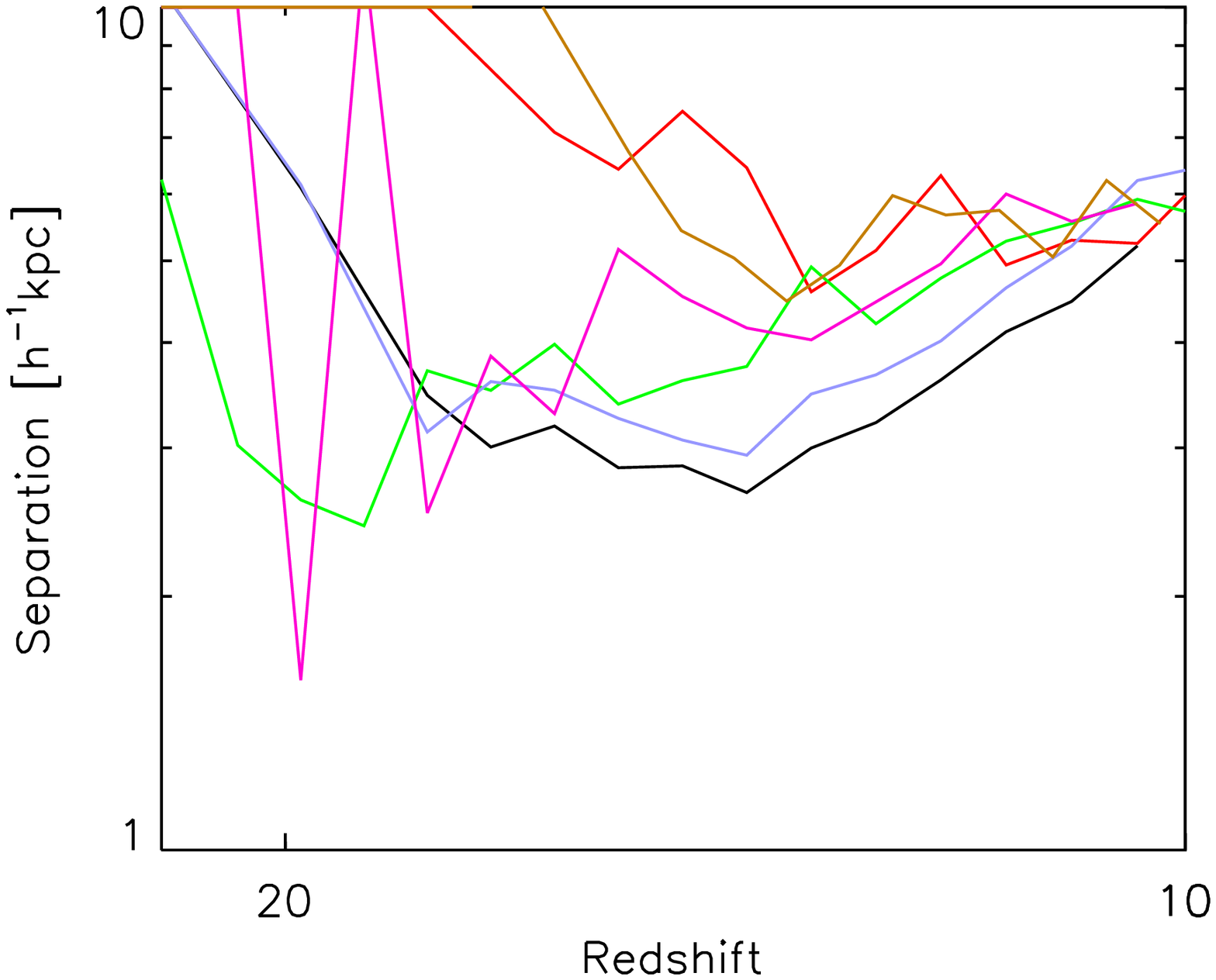}}
\caption{Clustering of \lq first galaxy\rq\ progenitor haloes 
  (virial temperature, $T_{\rm vir}>10^4$~K). {\em Left Panel}: mean
  number of progenitor halo pairs in Aq-A, as a function of proper
  separation, at the redshifts indicated in the panel. {\em Right
    Panel:} redshift evolution of the mean separation between pairs of
  haloes with $T_{\rm vir}>10^4$~K. By $z=10$, the mean separation
  between first galaxy haloes is $\sim 6$~\kpch\, and each halo has on
  average another 10 haloes within   $\sim 20$~\kpch. The typical
  distance to a neighbouring halo has a large scatter amongst
  realizations, reflecting the fact that these haloes only just began
  to form at these early times.}
\label{fig:galaxycor}
\end{figure*}

Progenitors of {\em Aquarius} haloes first achieve a virial
temperature, $T_{\rm vir}>10^4$~K, around $z\sim 25$; the abundance of
such ``first galaxy'' halos rises rapidly to a few tens by $z\sim 18$
and reaches a broad peak of $\sim 300$ by $z=10$, as shown in
Fig.~\ref{fig:firstgalaxyabundance}. The scatter amongst the six {\em
Aquarius} haloes is large. The rise in abundance is reasonably well
described by the Press-Schechter formula, but this overestimates the
number at $z=10$ by $\sim 50$ percent. As was the case for Pop.~III
haloes discussed earlier, the first galaxy haloes in a proto-Milky Way
object form at $z\sim 25$, substantially later than in a proto-cluster
region where they form at $z \sim 40$ (Gao et al. 2004a). The more
massive $z=0$ {\em Aquarius} haloes tend to have more first galaxy
progenitors at $z\sim 10$.\\

As before, we characterise the clustering of haloes that can cool
through atomic processes by the number of neighbours, $N(r)$, at
distance $r$. As shown in Fig.~\ref{fig:galaxycor}, $N(r)$ is again
approximately a power-law, $N(r) \propto r^2$. At $z=10$, a first
galaxy halo has a neighbouring first galaxy halo within $\sim
6$~\kpch\ and $\sim 10$ within $\sim 25$~\kpch. The halo-to-halo
scatter in clustering properties is large at early redshift $z\sim 20$
(Fig.~\ref{fig:galaxycor}, right panel), but as more haloes form, it
decreases; the typical separation reaches $r\sim 6$~\kpch\ at $z=10$.
Consequently, the ionised regions formed by star formation around such
first galaxies need not be very large, of order a few
\kpch\, before HII regions percolate across a proto-Milky Way galaxy
by $z=10$. \\

\begin{figure}
\resizebox{8cm}{!}{\includegraphics{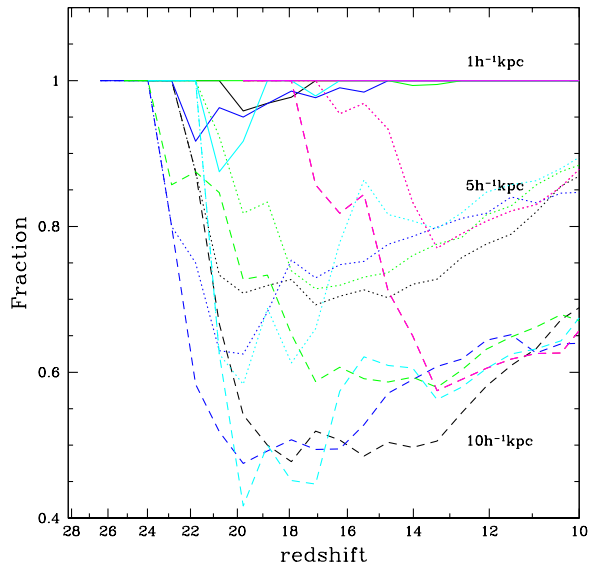}}
\caption{Fraction of first galaxies {\em not} 
affected by neighbouring first galaxies as a function of redshift,
if the radius of influence of the feedback is $R=$10, 5 or 1 \kpch\
(dashed, dotted, and solid lines, respectively). For the smaller
value of $R$, feedback has little or no effect, but for larger $R$,
feedback may decrease the number of first galaxies considerably, up
to a factor $\sim 2$.}
\label{fig:reion}
\end{figure}

Feedback from star formation in a first galaxy may potentially affect
a neighbouring halo. Figure~\ref{fig:reion} displays the fraction of
halos that would be affected depending on the value of the radius,
$R$, of influence of the feedback, when taking clustering into
account. For $R=1$\kpch, feedback has no effect, but for $R=5$~\kpch,
20-30 percent of potential first galaxy halos could be affected, rising to
40-50 percent for $R=10$\kpch.

\section{Relics of early star formation}
\subsection{Relics of the first stars}
\begin{figure}
\resizebox{8cm}{!}{\includegraphics{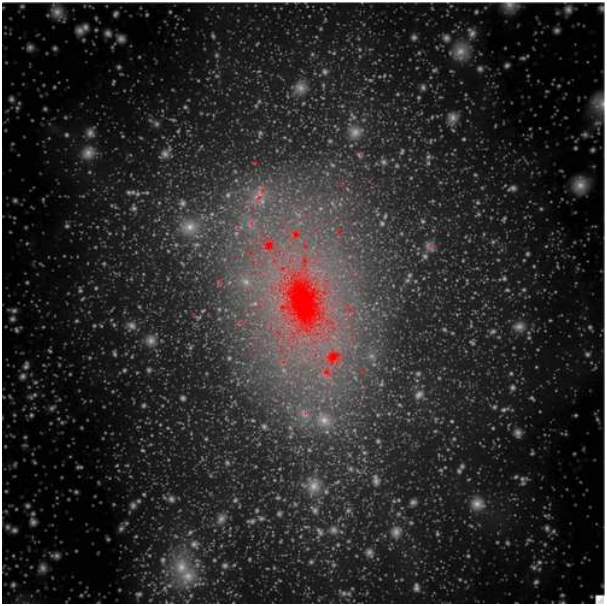}} 
\caption{Projected distribution at $z=0$ of first star relics
(red), compared to that of the dark matter for the Aq-A-2 halo. A
significant fraction of the stars lie in the central part of the main
galaxy, but some end up in dark matter subhalos. The image is 1080~kpc
on a side, approximately twice the diameter of the virialised dark
matter halo.}
\label{fig:stars-today}
\end{figure}
\begin{figure}
\resizebox{8cm}{!}{\includegraphics{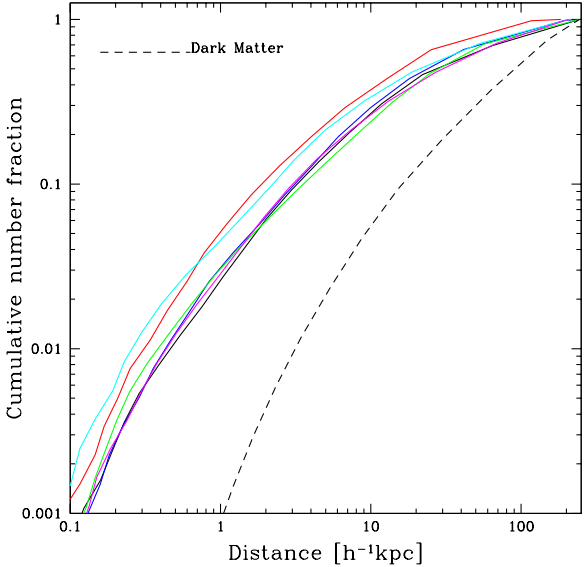}} 
\caption{Cumulative distribution of second generation, first star relics,
  stars at redshift $z=0$ for all six {\em Aquarius} haloes (colour
  lines), compared to the dark matter profile for the Aq-A halo
  (dashed black line).} 
\label{fig:stars}
\end{figure}

If, as seems likely, the first stars that form in a CDM universe are
massive, then only their dark remnants will remain in the Milky Way
today. However, evidence for their existence may be hidden in the
anomalous abundance patterns of long-lived second generation stars that
formed nearby, in gas predominantly enriched by Pop.~III stars
(e.g. Beers \& Christlieb 2005; Rollinde et al 2008); we will call
such stars \lq first star relics \rq. Where are these fossils of primordial
star formation to be found today? To answer this question, we make
use of the merger history of the haloes hosting primordial stars which
is readily available in the simulations. 

It seems reasonable to assume that feedback from Pop.~III stars will have
prevented the gas enriched by them from forming further stars until it
is accreted later into a first galaxy halo where it can cool by
atomic processes (see e.g. the simulations of Wise \& Abel 2008). The
earliest generation of stars formed in such a first galaxy may then
have the peculiar abundance pattern characteristic of Pop.~III
nucleosynthesis. However stars that form later may have been
enriched by the elements produced by the previous generation of
relatively metal rich stars, so that eventually the net abundance
pattern loses its characteristic Pop.~III star signature. To implement
this idea we proceed as follows. We assume that each first galaxy
halo contains a fixed {\em total mass} of {\bf \lq first star
  relics\rq} - {\bf stars that form in an environment enriched solely
  by truly primordial stars}  -- independent of redshift. To trace the
location of such stars at later times, we tag the $N=100$ most bound
particles {\bf at the time the virial temperature of the halo first
  exceeds $T=10^4$~K} \footnote{We verified that our final results do
  not depend strongly on the arbitrary  choice of $N$.}. This
  assumes that these second-generation stars form in the centres of
  such dark-matter haloes. Once a halo has started forming stars, it
will grow in mass, and may merge with other haloes. The number of
first star relics in a halo will only change if it merges with a halo
that contains its own population of first star relic stars. We
continue tagging first star relics in haloes that reach the threshold
$T_{\rm vir}$ until $z=10$, after which we assume that reionisation of
the Universe evaporates all mini-haloes.

In summary, we identify the epoch when a halo is first capable of
sustaining gas that can cool through atomic processes and tag its 100
most bound particles as the first star relics, until $z=10$. We
can then investigate the spatial distribution of these first star
fossils at any later epoch by tracing the tagged particles.  The
location of the first star relics today is illustrated in
Fig.~\ref{fig:stars-today} where, for clarity, we show only a
fraction of the stars. The image shows that most of the first
  star relics end up in the central regions of the halo, but some
survive inside subhalos.

The visual impression gained from Fig.~\ref{fig:stars-today} is
quantified in Fig.~\ref{fig:stars} which shows that the distribution
of the first star relics today (which is similar in all six {\em
  Aquarius} halos) is expected to be much more concentrated than that
of the dark matter. Fifty percent of these first star relics lie
within 20\kpch\ of the centre and $\sim 10$ per cent lie inside what
would be the galactic bulge, at $r<2\kpch$.  In contrast, only $\sim
10$ percent of the dark matter lies within 20\kpch, and only about 0.1
percent is at $r<1\kpch$. Even though the stars in our simulations are
already very concentrated, earlier models of the remnants of PoP-III
stars (White \& Springel 2000; Diemand et al. 2005) suggested they
should be even more concentrated, with almost all of them lying in the
bulge where they would be very hard to detect.

Although most of the first star relics reside in the central
parts of the main halo, a significant fraction end up in dark matter
subhalos. This is interesting because it may be easier to detect such
unusual stars in a nearby dwarf galaxy than to hunt for them in the
crowded surroundings of the Milky Way bulge.

\subsection{Relics of the first galaxies}

\begin{figure*}
\resizebox{8cm}{!}{\includegraphics{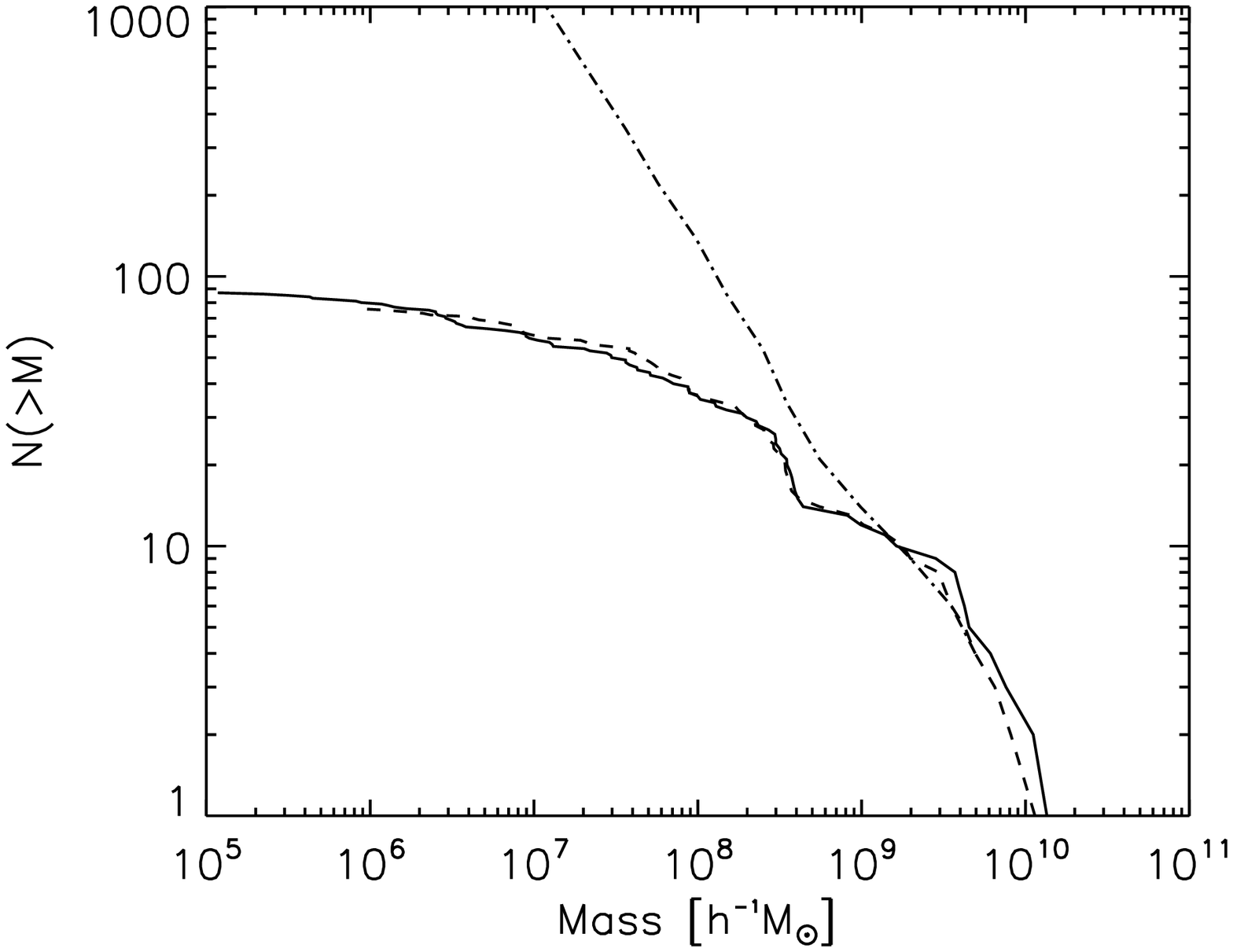}}
\resizebox{8cm}{!}{\includegraphics{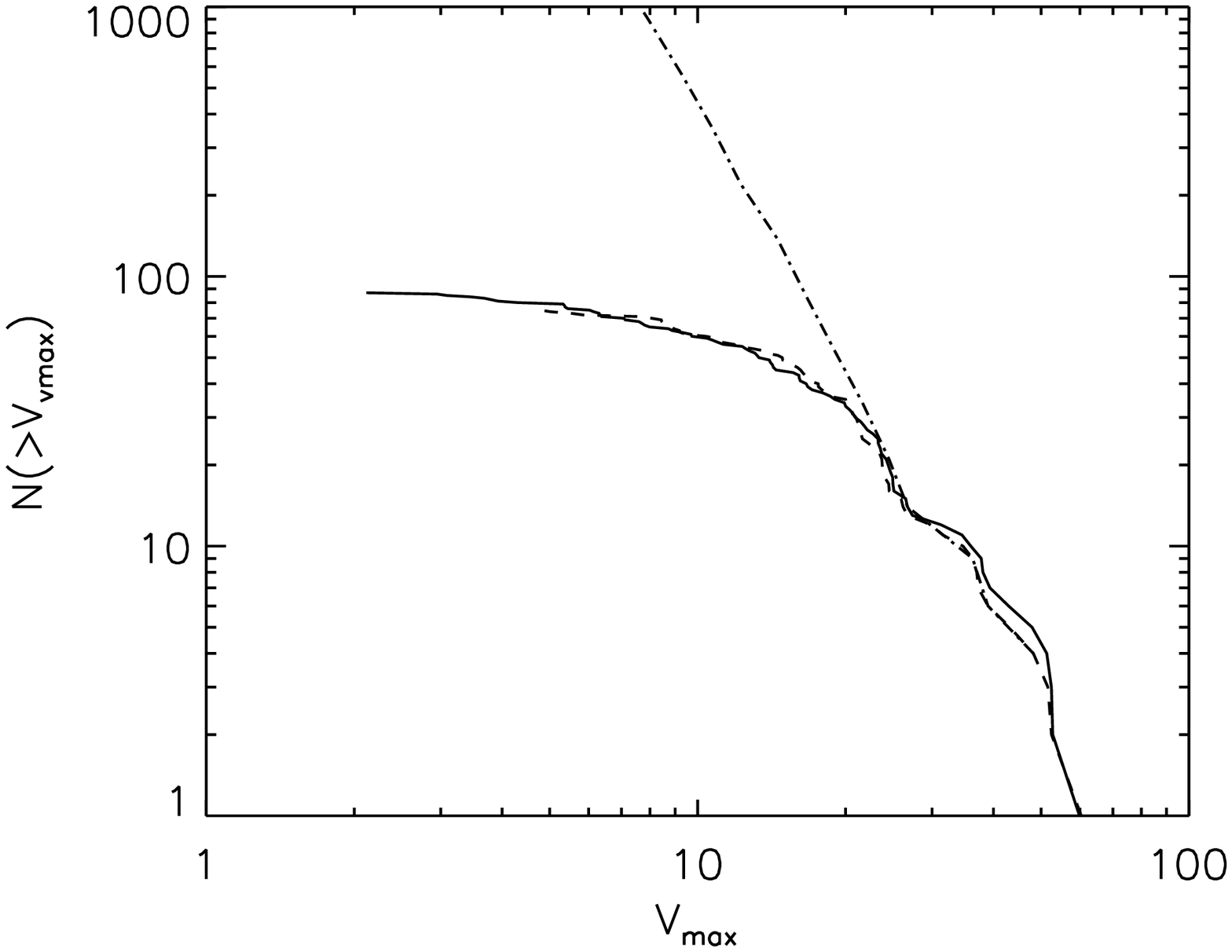}}
\caption{{\em Left panel}: Cumulative mass function at $z=0$ of 
`first galaxy' haloes (i.e. bound substructures which had a progenitor
at $z>10$ with $T_{\rm vir}>10^4$~K), for the Aq-A-1 (solid line)
and the 8$\times$ lower resolution Aq-A-2 (dashed line)
simulations. Numerical convergence is very good above 100 particles
(corresponding to $10^6\solarmass$ in Aq-A-2). For comparison, the
dashed-dotted line is the $z=0$ mass function of all gravitationally
bound substructures in Aq-A-1. {\em Right panel:} as the left panel,
but for the maximum of the circular velocity curve, $V_{\rm max}$.}
\label{fig:C02fateabundance}
\end{figure*}

In this section we investigate the fate of first galaxy halos. As
discussed in Section~\ref{sect:firstgals}, these are dark matter halos
which formed before $z=10$ and have $T_{\rm vir}>10^4$~K. We are
particularly interested in descendants which today are to be found
within the virial radius of the parent halo. To establish the range of
descendant masses which are reliably simulated, we begin by carrying
out a convergence test of the present-day mass function.

We compare results from the 1.4 billion particle Aq-A-1 simulation
with those from the Aq-A-2 simulation which has 8 times poorer
resolution. (All six {\em Aquarius} haloes have been simulated at
the resolution of the Aq-A-2 halo; see Table~\ref{TabSims}). The
cumulative mass function of first galaxy haloes that survive to $z=0$,
shown in Fig.~\ref{fig:C02fateabundance}, shows excellent numerical
convergence above a mass of $10^6\solarmass$, corresponding to 100
particles in Aq-A-2, and to a maximum circular velocity of $V_{\rm
max} \sim 5\kms$. We therefore adopt $N_p=100$ as the resolution limit
of the simulations. 

Aq-A-2 has 76 surviving first galaxy haloes with $N_p>100$
($M>10^6\solarmass$). Aq-A-1 has 88 survivors with $N_p>100$,
corresponding to $M>10^5\solarmass$ and $V_{\rm max}\sim 2\kms$. Thus,
Aq-A-1 resolves dark matter subhaloes with circular velocities lower
than the stellar velocity dispersions of the smallest dwarf galaxy
satellites of the Milk Way discovered to date (e.g. Simon \& Geha
2007).

\begin{figure*}
\resizebox{16cm}{!}{\includegraphics{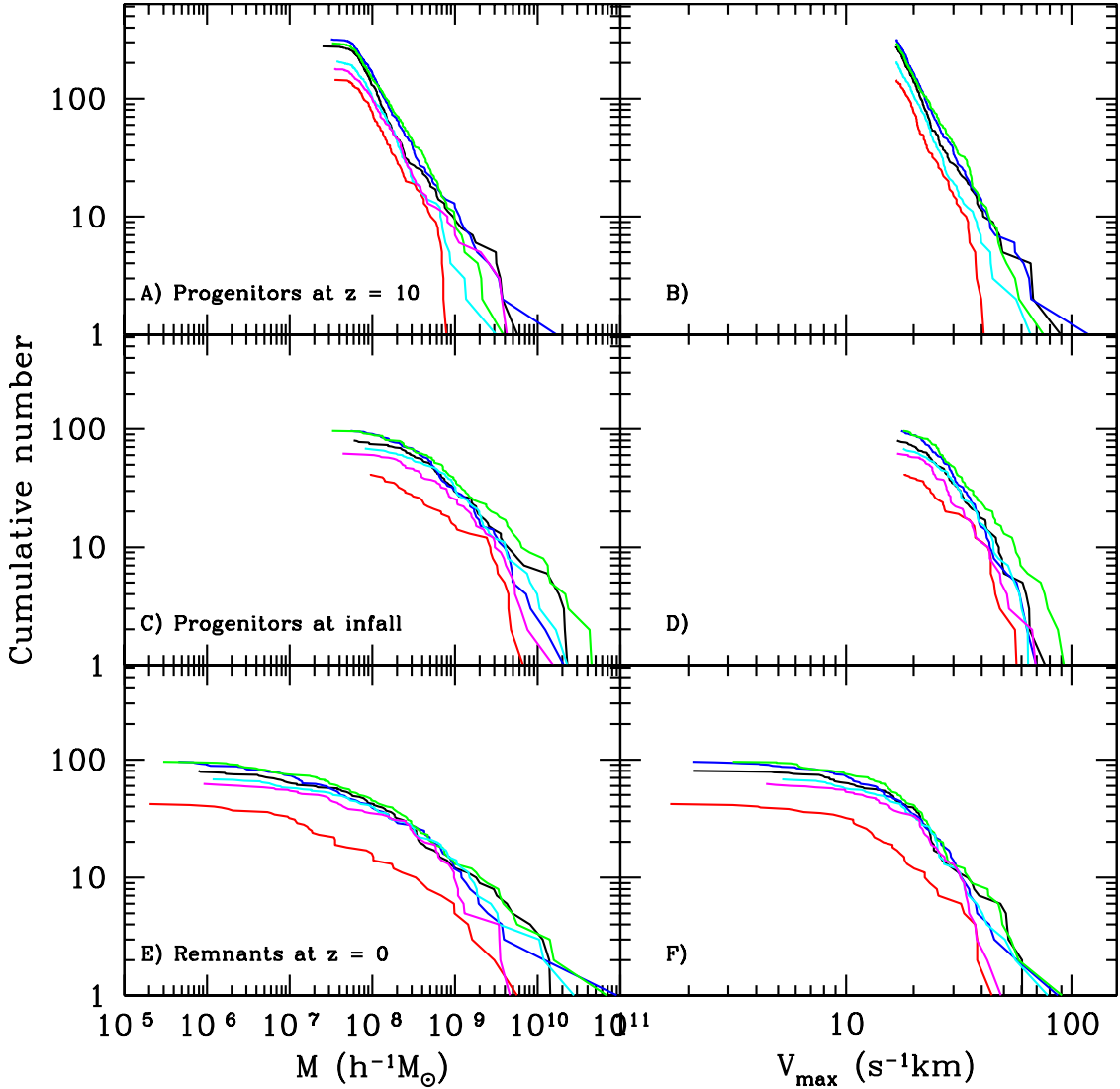}}
\caption{Statistics of first galaxy haloes in all six {\em Aquarius
haloes} (the colour scheme is as in Fig.~1) at z=10 (top), when they
are accreted by the main halo (middle), and at $z=0$ (bottom). The
left panels show cumulative mass functions and the right panels show
cumulative $V_{\rm max}$ distributions. There is a large variance
between the different {\em Aquarius} haloes at all redshifts. For
example, the mass of the most massive $z=0$ remnant varies between
$3\times 10^9\solarmass$ to nearly $10^{11}\solarmass$. The more
massive {\em Aquarius haloes} tend to have the more massive
substructures. Comparing the $z=0$ masses with those at infall or at
$z=10$ demonstrates that most substructures lose a considerable
fraction of their mass by the present.}
\label{fig:fateall}
\end{figure*}

\begin{figure}
\resizebox{8cm}{!}{\includegraphics{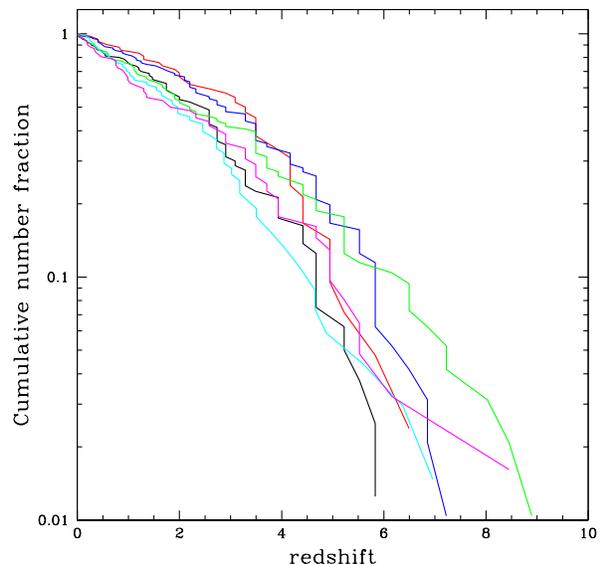}}
\caption{Cumulative fraction of first galaxy haloes that were accreted before a
given redshift and survive until $z=0$. The different lines correspond
to the 6 {\em Aquarius} halos. Typically 50~percent of haloes fell in
before $z=3$.}
\label{fig:history}
\end{figure}

\subsubsection{Abundance of the first galaxies and their descendants}

The mass function at $z=10$ of those \lq first galaxy\rq\ haloes which
ended up inside the virial radius ($\sim 170$\kpch\, see
Table~\ref{TabSims}) of the main halo today, and the corresponding
maximum circular velocity function, are shown in the top two panels of
Fig.~\ref{fig:fateall} for all {\em Aquarius} halos. The most massive
of these subhaloes has $M\sim 10^{10}h^{-1}M_\odot$, and circular
velocity, $V_{\rm  max}\sim 100$~km~s$^{-1}$. There are typically
$\sim 10$ subhaloes with $M>10^9\solarmass$ (corresponding to $V_{\rm
  max}>40\kms$) and 100 with $M>10^8\solarmass$ (corresponding to
$V_{\rm max}>20\kms$).

The lower panels of Fig.~\ref{fig:fateall} give the present day mass
and $V_{\rm max}$ distributions of the descendants of the first galaxy
halos illustrated in the top panels.  Every halo contains $\sim 80$
first galaxy descendants with mass $M>10^{6}\solarmass$ (corresponding
to $V_{\rm max}>5\kms$) that survived since $z=10$.  Note that there
is a large variation in the number of survivors in the different {\em
Aquarius} simulations, with the more massive haloes harbouring more
massive substructures on average. The mass of the most massive $z=0$
remnant varies from $3\times 10^9\solarmass$ to nearly
$10^{11}\solarmass$.

The lower right panel in Fig.~\ref{fig:fateall} shows that, at
$z=0$, the Aq-A-2 halo contains over 60 first galaxy remnant
subhaloes with $V_{\rm max}> 10$\kms inside its virial radius. This
is a factor of a few larger than the number of classical satellites
of the Milky Way. However the {\em total} (i.e. not just descendants
of first galaxies) number of subhalos with $V_{\rm max}>10\kms$ at
$z=0$ is $\sim 400$, as may be seen from the right panel of
Fig.~\ref{fig:C02fateabundance}. By contrast, for $V_{\rm max}
>22\kms$ (corresponding to $M\gsim 3\times 10^8\solarmass$), the
number of {\em surviving} first galaxy haloes equals the total
number of haloes. From SPH simulations of {\em Aquarius} halos,
Okamoto \& Frenk (2009) found that {\em all} subhalos with $V_{\rm
max}=23.5\kms$ host a satellite with V-band luminosity, $L_V> 2.5
\times 10^5 \solarl$, whereas below that value a diminishing
fraction of subhalos manage to make a visible satellite. It is
therefore tempting to identify these more massive first galaxy
remnants with bright satellite galaxies and to conclude that
these must have formed before $z=10$.

The stellar content of a present day subhalo is thought to be more
tightly connected to its dark matter mass at the time when it was
first accreted into the main halo than to its current dark matter mass
(e.g. Gao et al. 2004a,b. Libeskind et al. 2005). This is because
substructures may lose a significant fraction of their dark matter
once they have been accreted into the main halo, whereas the more
tightly bound stellar content is stripped to a lesser
extent. Fig.~\ref{fig:history} shows the redshift when first galaxy
halos surviving to the present were accreted by the main halo. The
median accretion redshift varies between 2 and 3.5 in the different
{\em Aquarius} halos. Less than 1 percent of the halos were accreted
before redshift 9 and less than 10 percent before redshift 7. Ninety
percent fell in before $z=1$.

The mass and maximum velocity functions at the time of accretion are
shown in the middle panels of Fig.~\ref{fig:fateall}. Approximately
half of the surviving substructures had a mass, $M> 5 \times
10^{8}\solarmass$ ($V_{\rm max}>30\kms$) at infall, compared to their
present day values of $M> 1 \times 10^{8}\solarmass$ ($V_{\rm
max}>15\kms$). Therefore such haloes have lost typically 80 per cent
of their mass, and their circular velocity has correspondingly
decreased by a factor $\sim 2$.

As discussed earlier, the sites of first galaxy formation are highly
clustered and, as a result, feedback may have had an important effect
on the number of first galaxies at $z=10$. However, it is likely that
feedback will have not affected the number of first galaxy {\em
remnants} at $z=0$ since it only influences nearby haloes which are
eventually likely to merge. So, although the {\em stellar properties}
of these substructures may have been affected by feedback at $z>10$,
the {\em number} of remnants is unlikely to depend on the details of
this process.  As an example, our most extreme feedback model
suppresses all star formation within 10\kpch\ from a first galaxy,
thereby decreasing the number of sites by 30 percent. Yet, the number
of first galaxy remnants in this model is the same as that of the
no-feedback model.

\begin{figure*}
\resizebox{16cm}{!}{\includegraphics{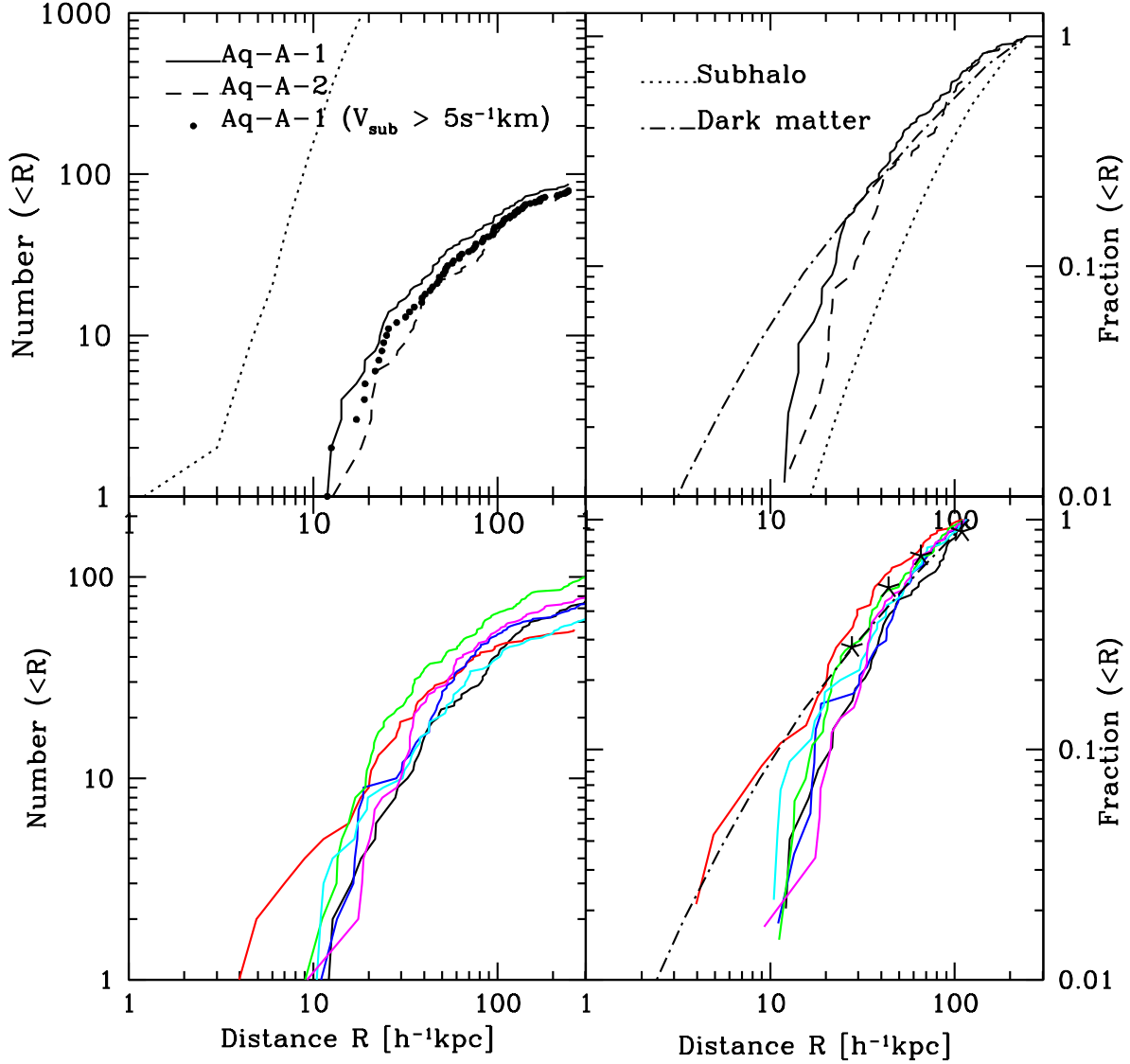}}
\caption{{\em Left panels}: present day cumulative number of first galaxy
halo remnants (i.e. substructures whose progenitors had $T_{\rm
vir}>10^4$~K at $z=10$), as a function of distance, $R$, from the
centre of the main halo. In the {\em top panel}, the solid line shows
all halos in Aq-A-1, the thick dotted line halos with $V_{\rm max}>5\kms$ in
Aq-A-1 and the dashed lines all halos in Aq-A-2. Halos with
$V_{\rm max}<5\kms$ are not resolved in Aq-A-2; for $V_{\rm max}>5\kms$ the
distributions in the two simulations agree within the noise. The thin
dotted lines shows the cumulative number of {\em all} substructures
with $V_{\rm max}>2\kms$ which is almost two orders of magnitude higher than
the number of first galaxy remnants.  The {\em bottom panel} shows
results for all 5 level-2 halos. {\em Top right panel}: as the left
panels, but showing the cumulative {\em fraction} normalised to the
number within the virial radius, for Aq-A-1 (thin solid line) and Aq-A-2
(dashed line). {\em Bottom right panel}: as the left panels, but
showing the cumulative {\em fraction} within $R=120\kpch$ for all 5
{\em Aquarius} halos. The black solid lines in the top and bottom
panels indicate the dark matter profile and the dotted line in the top
panel the radial profile of {\em all} substructures. The pentagons in
the lower panel correspond to known Milky Way satellites. Their radial
distribution resembles that of the outer first galaxy remnants.}
\label{fig:c02spa}
\end{figure*}

\begin{figure*}
\resizebox{8cm}{!}{\includegraphics{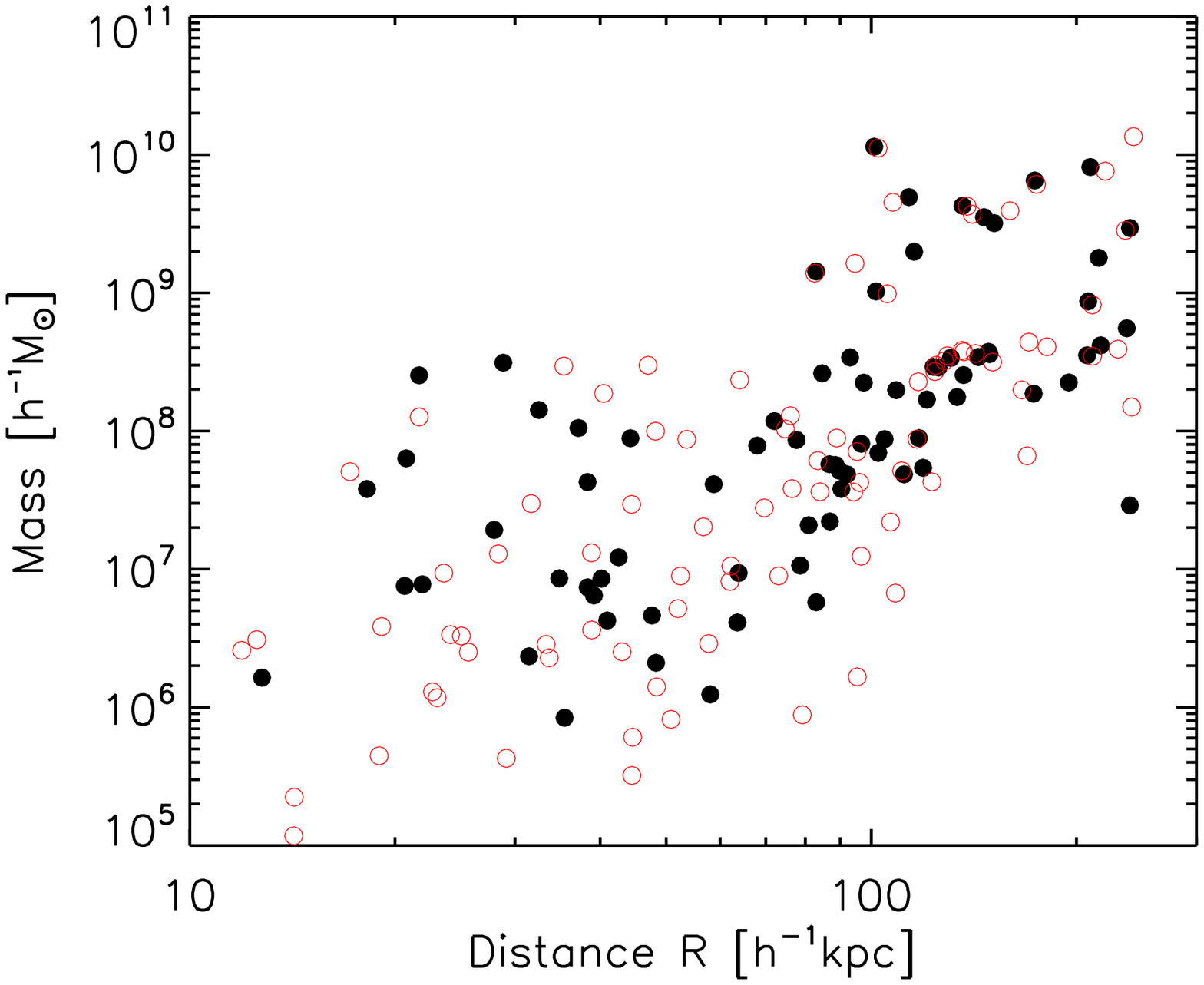}}
\resizebox{8cm}{!}{\includegraphics{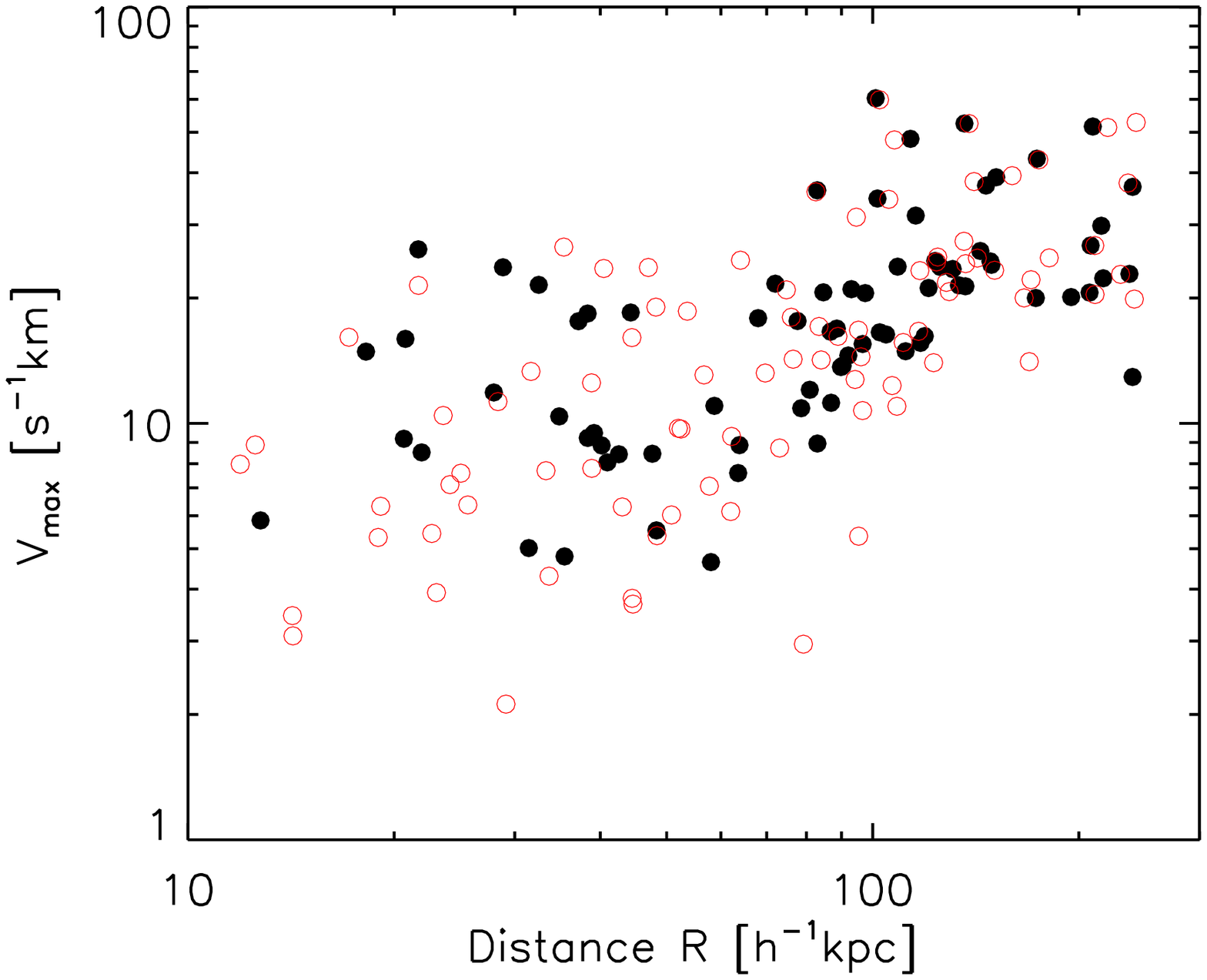}}
\resizebox{8cm}{!}{\includegraphics{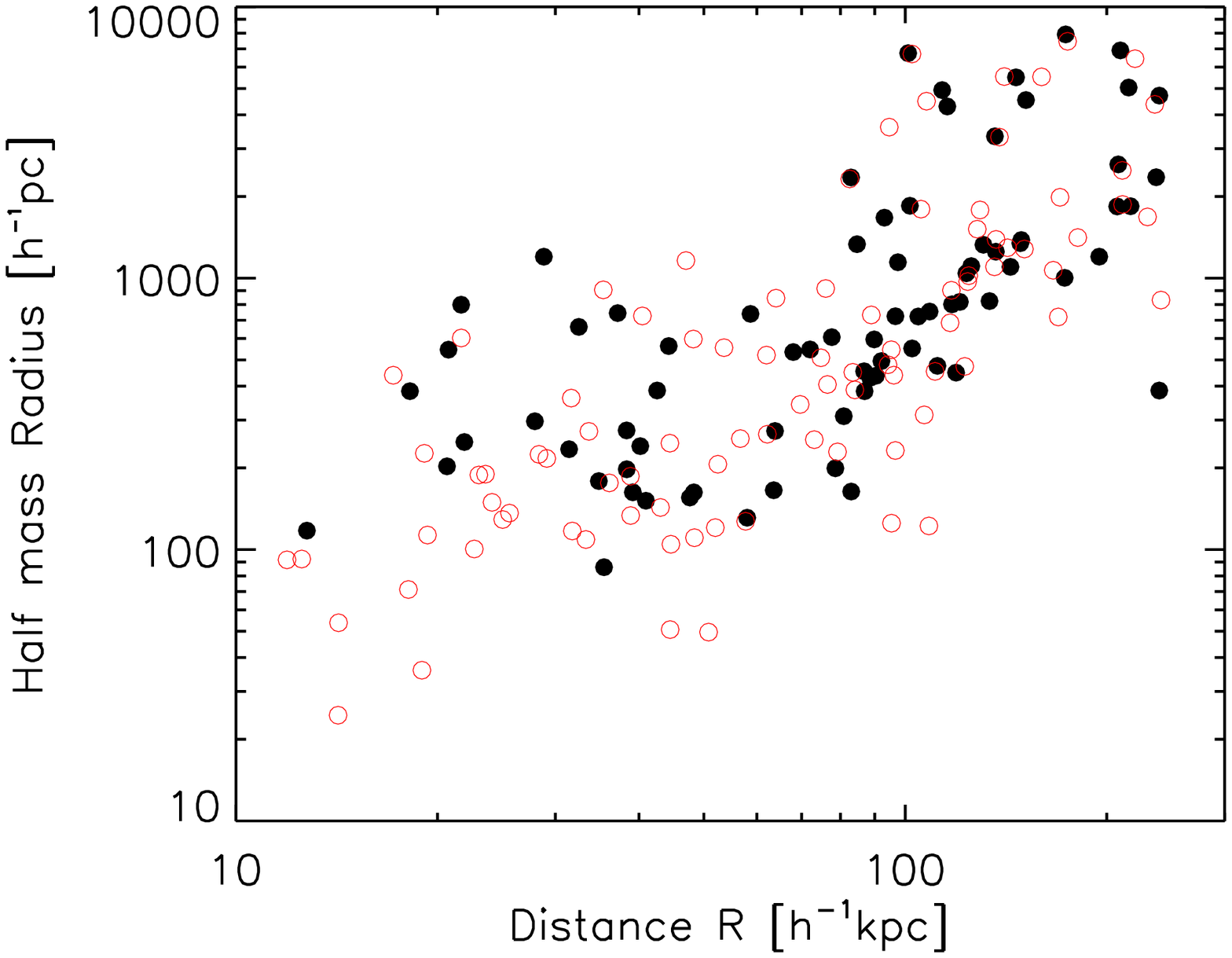}}
\resizebox{8cm}{!}{\includegraphics{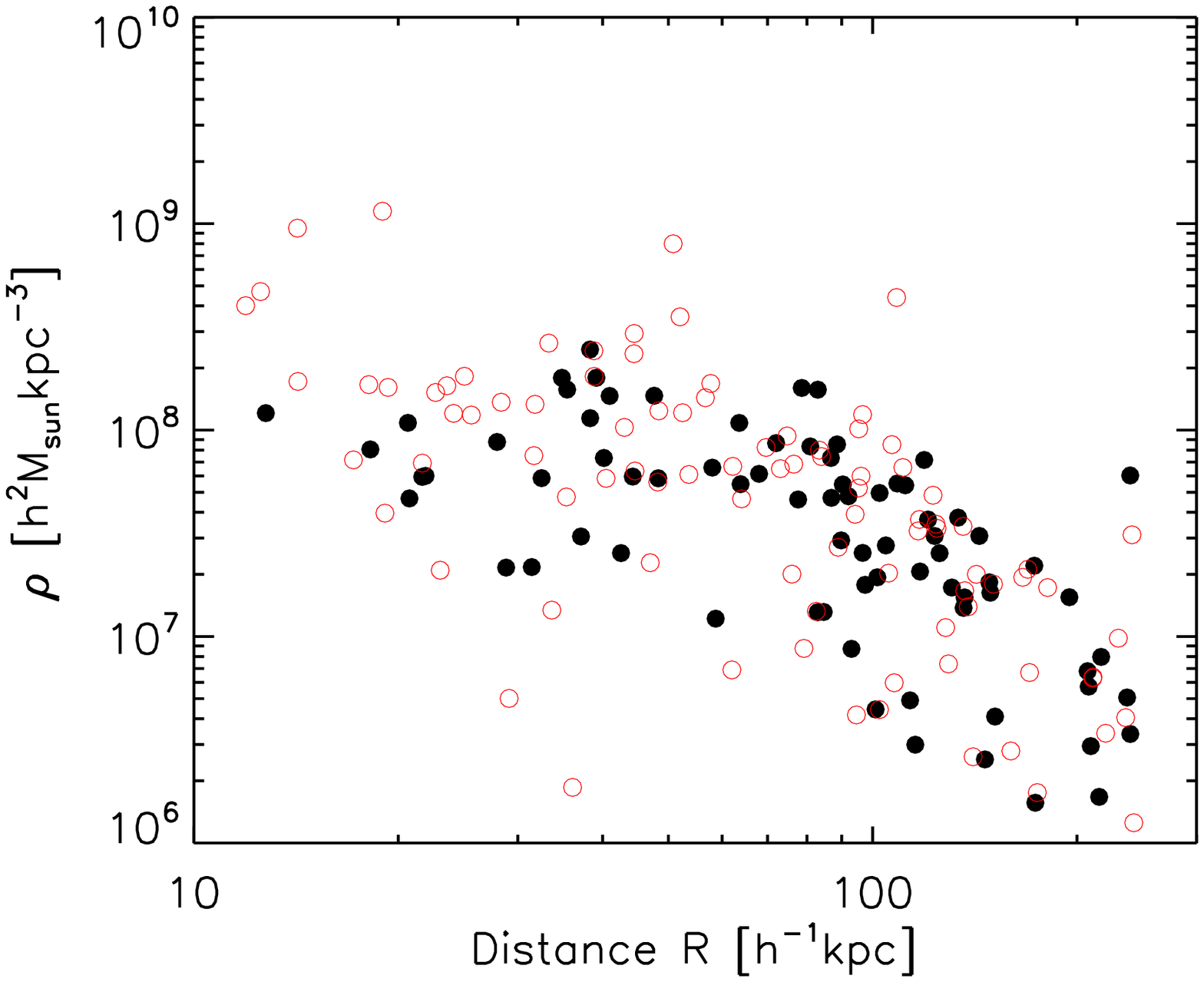}}
\caption{Mass, maximum circular velocity, $V_{\rm max}$, half-mass
  radius, and mean density within the half-mass radius, $\rho$, at
  $z=0$ of first galaxy haloes as function of their distance to the
  centre of the main halo: high resolution simulation Aq-A-1 ({\em
    open circles}), and the lower resolution simulation of the same
  halo, Aq-A-2 ({\em filled circles}). Resolution effects are
  noticeable inside $\sim 40\kpch$ where there are more haloes at
  higher resolution. More massive haloes tend to be located at $r
  >100\kpch$. At smaller $R$ there is a large scatter in halo
  properties, but a notable tendency towards increased subhalo
  concentration.}

\label{fig:c02sca}
\end{figure*}

\subsection{Spatial distribution of first galaxy remnants}

The remnants of the first galaxies are a small minority amongst
surviving subhalos. For example, of the $\sim 10^3$ surviving
subhaloes within $R=20\kpch$ in Aq-A-1, only $\sim 10$ are first
galaxy remnants. Their predicted spatial distribution is a useful
pointer as to whether observed halo objects could, in fact, be first
galaxy remnants.

The top left panel of Fig.~\ref{fig:c02spa} shows the cumulative
number of first galaxy remnants as a function of distance in both
Aq-A-1 and Aq-A-2. Both simulations have about 100 remnants within the
virial radius but in the inner part of the halo there are more
remnants in Aq-A-1 than in Aq-A-2. The deficit in Aq-A-2 is a
resolution effect. As may be seen in Fig.~ \ref{fig:C02fateabundance},
the smallest resolved halos in Aq-A-2 have $V_{\rm max}=5\kms$ whereas
in Aq-A-1 they have $V_{\rm max}=2\kms$. These small halos are missing
from the central parts of Aq-A-2. Indeed, for $V_{\rm max}>5\kms$, the
distributions in the two simulations agree within the errors, as may
be seen by comparing the dashed and dotted lines.

The bottom left panel in Fig.~\ref{fig:c02spa} shows that the
abundance of remnants at a given radius varies by about a factor of 2
between realizations. Five of them have no first galaxy
remnants closer to the centre than $10\kpch$, but one of them (Aq-B), has
5, including one at only $3.5\kpch$ from the centre. The top right
panel of Fig.~\ref{fig:c02spa} shows that the distribution of first
galaxy remnants is more centrally concentrated than the overall
distribution of {\em all} surviving subhaloes with $V_{\rm max}>5\kms$. For
example, 50\% of the first galaxy remnants are within $50\kpch$ of the
centre, whereas 50\% of the subhalo population is within twice that
distance. In the outer parts of the halo ($R\gsim 40\kpch$), the
distribution of first galaxy remnants tracks that of the dark matter
but, in the inner parts ($R\lsim 30\kpch$) it is shallower.

Typically, around 30 first galaxy halo remnants end up in the inner
35\kpch\ of the halo. In the Milky Way there are 6 known satellites
within this distance which, when allowing for the partial coverage of
observational searches for satellites, yields an expected number of
$\sim 25$ dwarfs (Tollerud et al. 2008). The cumulative radial
distribution of the 24 known Milky way satellites, normalised to the
total number within 120\kpch, is plotted in the lower right panel of
Fig.~\ref{fig:c02spa} and is interestingly similar to the distribution
of the first galaxy remnants. Detailed treatments of the satellite
populations in the {\em Aquarius} halos may be found in Cooper et
al. (2009), Okamoto et al. (2009) and Okamoto \&
Frenk (2009). 

Various physical properties of the first galaxy remnants: mass,
$V_{\rm max}$, half-mass radius, and mean density, $\rho$, within the
half mass radius are plotted in Fig.~\ref{fig:c02sca} as a function of
distance from the centre in Aq-A-1 and Aq-A-2.  The numerical
convergence of these properties is very good. As pointed out by
Springel et al. (2008), and apparent in the figure, many of the
substructures at the two resolutions can be matched spatially. As
noted earlier, halos with $V_{\rm max}<5\kms$ are resolved in Aq-A-1,
but not in Aq-A-2. This limit approximately corresponds to a mass of
$10^6 \solarmass$. Above the resolution limit of Aq-A-2 the results
for the two simulations agree remarkably well.

Within $\sim 80\kpch$ from the centre, there is a large scatter
in the properties of the remnants. Further out, there is a clear
increase in mass and $V_{\rm max}$ with radius, reminiscent of the
trends found by Springel et al. (2009) for the population of subhalos
as a whole. The half-mass radius of the remnants increases
systematically with halocentric distance, from about 100 pc in the
inner parts to $10\kpch$ in the outer parts. The central dark matter
densities of the remnants are correspondingly larger in the inner
parts. (The apparent convergence of the central density in Aq-A-2 to a
constant value at small distances is purely a resolution effect.) The
smallest remnants have masses and $V_{\rm max}$ values similar to
those of globular clusters. 

\section{Summary \& Discussion}

We have used the set of six high resolution cosmological simulations
of galactic halos from the {\em Aquarius Project} (Springel et
al. 2009) to investigate the formation and fate of haloes capable of
hosting the first stars and the first galaxies. Our main results may
be summarized as follows:

\medskip
\noindent{\it First stars}

\begin{enumerate}

\item The first halos with virial temperature high enough to form 
 stars, according to the model of Gao et al. (2007), appear at
 redshift $z=35$; the number of such haloes increases exponentially to
 $\sim 2\times 10^3$ by $z=20$ and to $\sim 10^4$ by redshift $z=10$.

\item First star haloes are very highly clustered. The mean
  interhalo separation increases, in physical units, from
  $\sim0.5$\kpch\ at $z=25$ to $\sim 1$\kpch\ by $z=10$.  Thus, if the
  HII region created by a Pop.~III star is larger than a few hundred
  parsecs, then feedback due to ionising radiation is likely to have
  had a large impact on the number of first stars in the proto-galaxy.

\item The spatial distribution of the remnants of primordial stars today, as
  traced by the most strongly bound dark matter particles in the
  original halo, is more concentrated than that of the dark matter,
  and scales as $n(R) \sim R^{-2.9}$ with distance, $R$, from the
  galactic centre. A few percent of the remnants lie in the galactic
  bulge, $\sim 10$ percent in the central $10$\kpch\ of the halo, and
  $\sim 50$ percent in the inner $30$\kpch\ of the galactic halo. A
  significant fraction is associated with subhalos, suggesting that
  the remnants of some of the first stars could be found today in
  satellite galaxies, confirming earlier conjectures (e.g. Ricotti \&
  Gnedin 2005; Munoz et al. 2009).
\end{enumerate}

\medskip
\noindent{\em First galaxies}

\begin{enumerate}
\item The first pre-galactic halos capable of hosting gas that
  could undergo atomic hydrogen cooling appear at redshift $z \sim
  25$; their number increases to $\sim 100$ by redshift $z=20$ and to
  $\sim 300$ by at redshift $z=10$, with substantial scatter from halo
  to halo. The mean interhalo separation is $4$\kpch\ at redshift $z
  \sim 20$, and does not evolve much until $z=10$. Early
  pre-ionisation by UV photons produced in the first galaxies may have
  had a significant effect on their numbers at $z=10$, although the
  total number of remnants left over by $z=0$ is unaffected because
  most haloes that could plausibly suppress each other's star
  formation merge by $z=0$ anyway.

\item The majority of the surviving first galaxy halo remnants were
 accreted into the main halo early.  Fifty percent fell in before
 $z=3.5$ and ninety percent before $z=1$.

\item About 80 first galaxy halo remnants end up within the virial radius
 ($\sim 160$\kpch) of a galaxy like the Milky Way as satellites with
 dark matter mass above $10^6$\solarmass\ (corresponding to a maximum
 velocity $V_{\rm  max}\sim 5$\kms). More than half of them had a mass
 above $5 \times 10^{8}$\solarmass\ ($V_{\rm max}\sim 20$\kms) before
 they were  accreted by the main progenitor. Many are heavily stripped
 after infall.

\item The spatial distribution of the first galaxy halo remnants today
  follows that of the dark matter in the outer parts of the galactic
  halo, but is much less concentrated towards the centre. The
  remnants,  however, are significantly more concentrated than the
  substructure population as a whole. A typical {\em Aquarius} halo
  has 20 first galaxy remnants within $30$\kpch\ of the centre with a
  velocity dispersion of few kilometers per second, similar to the
  faintest Milky Way satellites known. Of course, the presence of
    the disk may have affected the survival of early substructures.

\item The mass, $V_{\rm max}$, and half-mass radius of the first
 galaxy remnant halos increase with radius. The largest, most massive
 halos are typically found beyond $100\kpch$ of the centre. The
 scatter in these trends increases substantially at smaller radii. At
 $r<100\kpch$, the median mass of the remnant subhaloes is $\sim
 10^7\solarmass$ and the median $V_{\rm max}$ is $\sim7\kms$.
\end{enumerate}

Our results suggest the following history of early star formation in
a galaxy like the Milky Way. Star formation begins at around $z=35$
in mini-halos and the number of sites where these Pop.III stars can
form rapidly builds up to thousands or even tens of thousands
depending on the efficiency with which energy from one star prevents
neighbouring haloes from making their own star. Since these
potential sites are strongly clustered, it is to be expected that
many of them will remain barren for some time. A second generation
of stars (first star relics) forming near the first stars may be
long-lived and their remnants would be found primarily in the inner
parts of the galaxy today but also inside satellite galaxies, where
they could be identified though their anomalous metallicity patterns.

Following this episode of metal-free star formation, the first
\lq galaxies\rq\ begin to form  around $z=25$. The presence of
local radiation sources, enrichment of gas by heavy elements, and
possibly mechanical feedback from the first supernovae, may all
contribute to a change in the stellar initial mass function, leading
to the formation of the first low mass stars. A significant fraction
of the haloes that hosted these first galaxies are destroyed by
mergers or stripping and their stellar remnants could have
contributed to the stellar halo (Cooper et al. 2009). Over 20
percent of the primordial fossils survive as bound structures until
the present and some could have developed into satellite
galaxies. Our simulations therefore suggest that some of the Milky
Way's dwarf satellites should contain a small fraction of the oldest
stars that formed during the early stages of formation of the Milky
Way. 
\section*{acknowledgments}

The {\it Aquarius} simulations were carried out as part of the
programme of the Virgo Consortium on the Leibniz Computing Centre in
Garching, the Cosmology Machine at the Institute for Computational
Cosmology in Durham and on the STELLA supercomputer at Groningen. LG
acknowledges support from an STFC Advanced Fellowship,  
one-hundred-talents program of the Chinese academy of science(CAS)
and the National basic research program of China (973 program under
grant No. 2009CB24901). CSF acknowledges a Royal Society Wolfson
Research Merit Award. This work was supported in part by an STFC
rolling grant to the ICC.

\end{document}